\documentclass[aps,prl,twocolumn,showpacs,showkeys,floatfix,nofootinbib]{revtex4-1}

\usepackage{amsmath}
\usepackage{amssymb}
\usepackage{graphicx}
\usepackage{relsize}
\usepackage{bm} 
\usepackage{color}

\renewcommand{\i}{\ensuremath{\mathrm{i}}}

\begin{document}
\title{Proximitized Materials}

\author{Igor \v{Z}uti\'c$^1$}
\email{zigor@buffalo.edu}
\author{Alex Matos-Abiague$^{2}$}
\author{Benedikt Scharf$^3$}
\author{Hanan Dery$^4$}
\author{Kirill Belashchenko$^5$}
\affiliation{
$^1$Department of Physics,  University at Buffalo, State University of New York, Buffalo,  New York 14260, USA \\
$^2$Department of Physics and Astronomy Wayne State University
Detroit, Michigan 48201, USA                                                                             \\
$^3$Institute for Theoretical Physics and Astrophysics, University of W\"{u}rzburg, Am Hubland, 97074 W\"{u}rzburg, Germany \\
$^4$Department of Electrical and Computer Engineering, and Department of Physics and Astronomy, University of Rochester, Rochester, New York 14627, USA \\
$^5$Department of Physics and Astronomy and Nebraska Center for Materials and Nanoscience, University of Nebraska-Lincoln, Lincoln, Nebraska 68588-0299, USA
}
\date{\today}

\begin{abstract}
Advances in scaling down heterostructures and having an improved interface quality together with atomically-thin two-dimensional 
materials suggest a novel approach to systematically design materials. A given material can be transformed through proximity effects 
whereby it acquires properties of its neighbors, for example, becoming superconducting, magnetic, topologically nontrivial, or with 
an enhanced spin-orbit coupling. Such proximity effects not only complement the conventional methods of designing materials by 
doping or functionalization, but can also overcome their various limitations. In proximitized materials it is possible to realize  properties 
that are not present in any constituent region of the considered heterostructure. While  the focus is on magnetic and spin-orbit proximity 
effects with their applications in spintronics, the outlined principles provide also a broader framework for  employing other proximity effects 
to tailor materials and realize novel phenomena.

\end{abstract}

\maketitle

\section{1. Introduction}\label{Sec:Intro}

Pristine materials seldom appear as we want them. Instead, their appeal typically comes from suitable modifications. The success of semiconductors  
is largely derived from doping where impurities are intentionally introduced to alter their properties. 
Doping is a critical part for a wide range of semiconductor applications, from transistors and solar cells, to light emitting diodes and lasers, recognized by multiple Nobel Awards~\cite{Yu:2010}. 
Beyond semiconductors, chemical doping is ubiquitous to many other materials and the resulting changes in chemical composition can produce striking results. 
Parent compounds of several copper-oxide layered materials at low doping are insulating antiferromagnets, at optimal doping high-temperature superconductors, 
and at high doping resemble conventional metals~\cite{Philips:2012}. 

\begin{figure*}
\vspace{-0.5cm}
\resizebox{13.5cm}{!}{\includegraphics{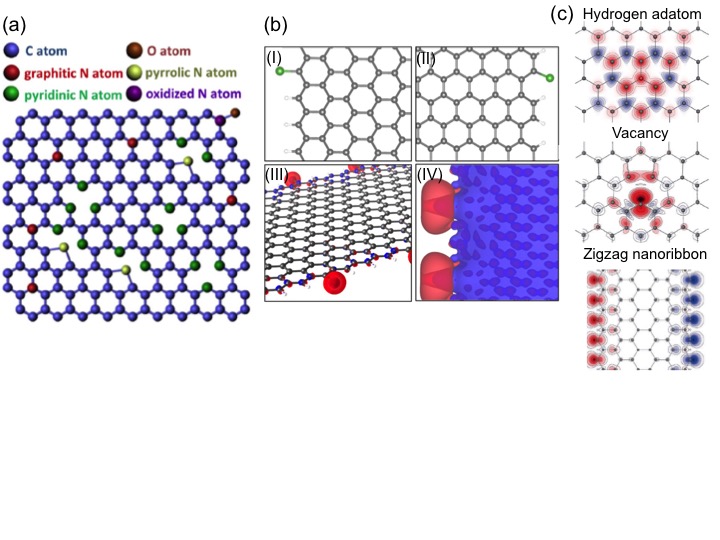}}
\vspace{-3cm}
\caption{\footnotesize{%
(a) Four types of nitrogen doping in graphene~\cite{Wei2009:NL}.
(b) Model of Cl-functionalized graphene zigzag (I) and armchair (II) nanoribbons (GNR). (III) Total charge density difference between Cl-functionalized and pristine zig-zag GNR. (IV) Electrostatic potential difference induced by the Cl functionalization for a zig-zag GNR. Charge density (potential) isosurface level: 0.05 e/\AA$^3$ (0.1 eV); positive isosurface values: blue, negative: red, C atoms: gray, Cl atoms: green, and H atoms: white~\cite{Quesnel2015:2DM}.
(c) Magnetic moment in graphene due to light adatoms and vacancy defects. 
Prediction of magnetic moments in graphene due to hydrogen, vacancy defects, and at the graphene edges. Red and blue: the opposite spin polarizations~\cite{Han2014:NN}. Adapted with permission (a) from Ref.~\cite{Wei2009:NL}, (b) from Ref.~\cite{Quesnel2015:2DM}, (c) from Ref.~\cite{Han2014:NN}.
}}
\label{fig:dope}
\vspace{-0.2truecm}
\end{figure*}

A common approach to improve a large class of low-dimensional materials is
by their chemical functionalization including chemical reactions with organic and inorganic molecules~\cite{Voiry2014:NC,Yan2012:CRC,Layek2015:CM,Puri2017:ACSN}. 
Several examples of doping and functionalization 
are illustrated in Fig.~\ref{fig:dope} for graphene, two-dimensional (2D) $sp^2$-hybridized carbon forming a honeycomb 
lattice~\cite{Quesnel2015:2DM,Wei2009:NL,Han2014:NN}.
Often the notion of functionalitization is extended to also include chemical changes induced by atoms, such as hydrogenated and flourinated 
graphene~\cite{Han2014:NN,Yazyev2007:PRB,McCreary2012:PRL,Gonzalez-Herrero2016:S}.

To understand some of the challenges in bringing about novel materials properties by doping, it is instructive to revisit the push to realize dilute magnetic semiconductors 
(DMS)~\cite{Furdyna1988:JAP,Dietl2014:RMP}.\footnote{{\bf Abbreviations:}
DMS dilute magnetic semiconductor, F ferromagnet, N nonmagnetic region, SOC spin-orbit coupling, SO spin orbit,
vdW van der Waals, TMD transition metal dichalcogenide, ML monolayer, DOS density of states,
FET  field effect transistor, MOS metal-oxide-semiconductor, MRAM magnetic random access memory, 
MR magnetoresistance, CB conduction band, VB valence band, RKKY Rutherman-Kittel-Kasuya-Yoshida,
IEC interlayer exchange coupling, 
TMR tunneling magnetoresistance, AF antiferromagnet, ISOC interfacial spin-orbit coupling, TAMR tunneling anisotropic magnetoresistance, 
AMR anisotropic magnetoresistance, CAMR crystalline anisotropic magnetoresistance, PIA pseudospin inversion asymmetry, 
LSV lateral spin valve, MLG magnetologic gate, and VCSEL vertical cavity surface emitting laser.}

Doping common semiconductors by magnetic impurities, typically Mn, was expected to realize in a single materials system a versatile control of charge degrees 
of freedom, characteristic for semiconductors, with the nonvolatile manipulation of spin and robust magnetism from ferromagnetic metals. Effectively, this could be a very 
desirable platform to implement a seamless integration of logic and memory.  The carrier-mediated magnetism in DMS  offers a control of the exchange interaction 
by tuning the Curie temperature, $T_C$,  through changes in the carrier density, by an applied electric field, photoexcitation, or even heating~\cite{Koshihara1997:PRL,Ohno2000:N,Petukhov2007:PRL,Zutic2004:RMP}, 
as well as reveal novel methods to control the direction of magnetization~\cite{Chernyshov2009:NP}.

The two most studied classes of Mn-doped  magnetic semiconductors are II-VI 
and III-V compounds~\cite{Furdyna1988:JAP,Dietl2014:RMP}. 
In the II-VI DMS Mn$^{2+}$ is isovalent with the group II ions and provides only spin doping; the lack of carriers makes robust ferromagnetism elusive, 
the $T_C$ is limited to a few K~\cite{Dietl2014:RMP}. In common III-V DMS, including the best studied example of (Ga,Mn)As, this leads to both spin and carrier doping, but a low solubility 
limit for Mn makes the growth very challenging and can lead to nanoscale clustering of Mn ions~\cite{DeBoeck1996:APL}. The presence of such nanoclusters often complicates 
an accurate 
determination of $T_C$ as well as of whether the compound is actually in a single phase~\cite{Zutic2004:RMP}.

However, even with a successful realization of a single phase DMS,  which for (Ga,Mn)As requires complex low-temperature molecular beam epitaxy,  the ferromagnetism 
is not supported at room temperature ($T_C \lesssim$ 190 K in (Ga,Mn)As~\cite{Dietl2014:RMP}), there are unintended materials changes. Excellent optical properties of GaAs, including 
strong luminescence,  are significantly diminished in (Ga,Mn)As, while with Mn-doping a low temperature mobility of GaAs that exceeds 1000 cm$^2$/V$\;$s, is reduced by 2-3 orders of magnitude. Similar limitations also pertain to functionalization,  known to result in disorder and significantly reduce the mobility of graphene. 
Graphene functionalization occurs randomly, posing a challenge to control how and where chemical reactions occur~\cite{Yan2012:CRC}.

\begin{figure}[ht]
\resizebox{8.8cm}{!}{\includegraphics{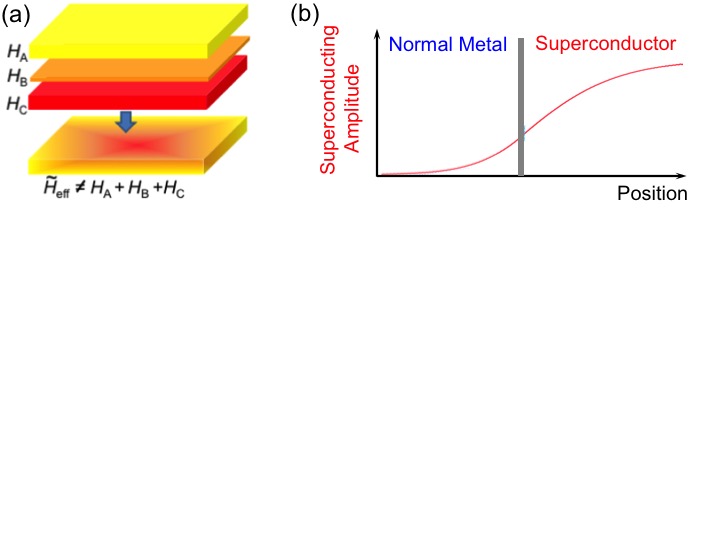}}
\vspace{-4.5truecm}
\caption{\footnotesize{%
(a) -H: Proximity modified layer  B in the presence of 
layers A, C, with the respective effective and individual Hamiltonians,
$\tilde{H}_\mathrm{eff}$, $H_A$, $H_B$, $H_C$. (b) Penetration of
superconductivity across an interface into a normal (nonsuperconducting) region.
Adapted with permission (b) from Ref.~\cite{Valls2010:PRB}.
}}
\label{fig:ABC}
\vspace{-0.2truecm}
\end{figure}

A radically different path to tailor materials has recently emerged from proximity effects which can transform a given material through 
its adjacent regions to become superconducting, magnetic, or topologically nontrivial. 
While proximity effects are commonly viewed 
as just curious and specialized phenomena limited to cryogenic temperatures or disappearing beyond a
few nanometers~\cite{deGennes1964:RMP,Hauser1969:PR,Buzdin2005:RMP}, in this review 
we elucidate a much broader picture of proximity effects as a ubiquitous approach to transform a wide class of materials that could 
overcome limitations inherent to doping and functionalization. Opportunities to design proximitized 
materials already arise at equilibrium as schematically illustrated in Fig.~\ref{fig:ABC}(a). The effective Hamiltonian describing a proximity-modified layer B, $\tilde{H}_\mathrm{eff}$, 
contains properties that are  different or absent from those in the individual regions, A, B, C.  

\begin{figure*}
\resizebox{17.5cm}{!}{\includegraphics{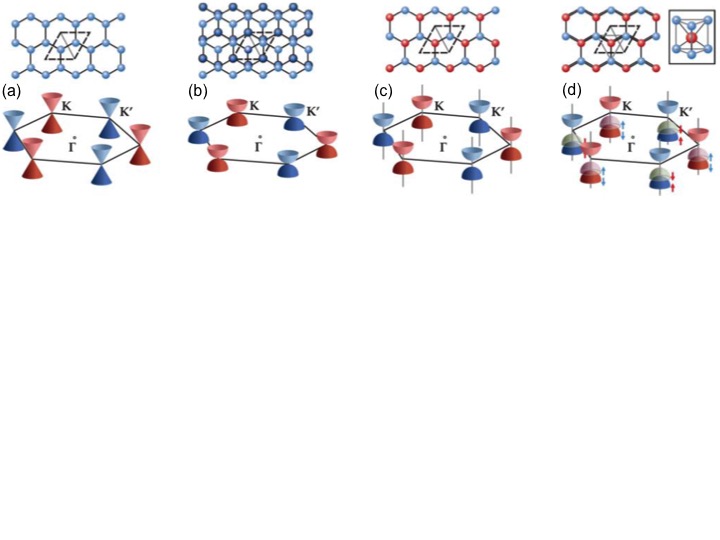}}
\vspace{-8.5truecm}
\caption{\footnotesize{%
vdW materials: lattice and band structures at the corners of the first Brillouin zone. (a) Monolayer (ML) graphene. (b) Bilayer graphene. (c) h-BN.
(d) Transition metal dichalcogenides (TMDs). The dashed lines  denote unit cells. Due to their inversion 
symmetry, ML and bilayer graphene have only a small bandgap~\cite{Han2014:NN}. With a large 
SOC,  the valence band in TMDs 
is split into two spin-polarized bands, marked by red and blue arrows. A smaller conduction band SOC is not shown. 
Spin reversal between the two valleys, $K$ and $K'$, reflects the spin-valley coupling. 
Adapted with permission from Ref.~\cite{Ajayan2016:PT}.}}
\label{fig:vdW}
\vspace{-0.2truecm}
\end{figure*}

The intuition about proximity effects is well-derived from the superconducting case, known for 85 years~\cite{Holm1932:ZP}. As shown in Fig.~\ref{fig:ABC}(b), 
superconducting properties can penetrate 
from a superconductor into a neighboring normal region which by itself would not be superconducting. 
Similarly, in magnetic proximity 
effects a magnetization 
from a ferromagnet (F) penetrates into a neighboring nonmagnetic region (N). Remarkably, superconducting proximity 
effects can attain orders of magnitude longer lengths than for other proximity effects,  even $>$ 100 $\mu$m in clean metals at sub-Kelvin 
temperatures~\cite{Wolf1978:RPP,Kompaniiets2014:JS}. 
Superconducting proximity can attributed to the process of Andreev reflection: at
an interface with a superconductor an incoming electron is retro-reflected as a hole, accompanied by a creation of a 
Cooper pair~\cite{Zutic2004:RMP}. While in a 
narrow sense proximity effects pertain to the transfer of an ordered state (i.e., superconductivity or magnetism) to another region where it 
was initially absent without strongly affecting its electronic structure, in recent years this term has been applied more broadly to also include 
proximity-induced spin-orbit coupling (SOC) 
or topological properties~\cite{Han2014:NN,Alicea2012:RPP}.   

In bulk materials, the sample size often largely exceeds the characteristic lengths of proximity effects allowing their neglect. However, in 
atomically-thin van der Waals (vdW) materials such as graphene, h-BN, and transition-metal dichalcogenides (TMDs)\cite{Geim2013:N,Ajayan2016:PT,Yazyev2015:MT,Wang2012:NN,Xu2014:NP} 
depicted in Fig.~\ref{fig:vdW}, the situation is drastically different, even short-range magnetic proximity effects exceed their thickness and strongly modify 
transport and optical properties.  
For example, pristine graphene is gapless and massless with a linear dispersion around the $K$  point 
in the Brillouin zone (Dirac cone), it has a negligible SOC and its density of states 
(DOS) is spin unpolarized. 
However, proximity effects from neighboring materials profoundly alter graphene's character such that it can acquire a positive or negative effective mass~\cite{Hunt2013:S}, spin polarization~\cite{Lazic2014:PRB}, SOC~\cite{Han2014:NN,Avsar2014:NC,Wang2015:NC,SaveroTorres2017:2DM},  
or even superconductivity~\cite{Efetov2015:NP,Heersche2007:N,Cao2018:Na}. 
Graphene is among many 
vdW materials that illustrate the emerging trends in tailoring their properties through proximity effects.
Furthermore, with a 
scaling-down of nanostructures and an improved quality of interfaces, other classes of materials are also becoming a suitable platform to demonstrate proximity effects.

While our review is mostly focused on magnetic and spin-orbit (SO) proximity effects and their applications to spintronics, 
the outlined framework for realizing proximitized materials provides also guidance to other intriguing opportunities. 
For example, proximity effects can be used to design exotic topological phases which reflect global properties of heterostructures 
insensitive to disorder and local perturbations, leading to applications such as ultra-high density magnetic 
storage using magnetic skyrmions~\cite{Fert2013:NN,Hrabec2017:NC,Romming2013:S,Hsu2017:NN}, 
or topologically protected quantum computing 
with non-Abelian quasiparticles~\cite{Aasen2016:PRX,DasSarma2015:QI,Kitaev2003:AP}.

\section{2. Electrostatic Gating with 2D Systems}\label{Sec:Gate} 
\begin{figure}[ht]
\resizebox{8.7cm}{!}{\includegraphics{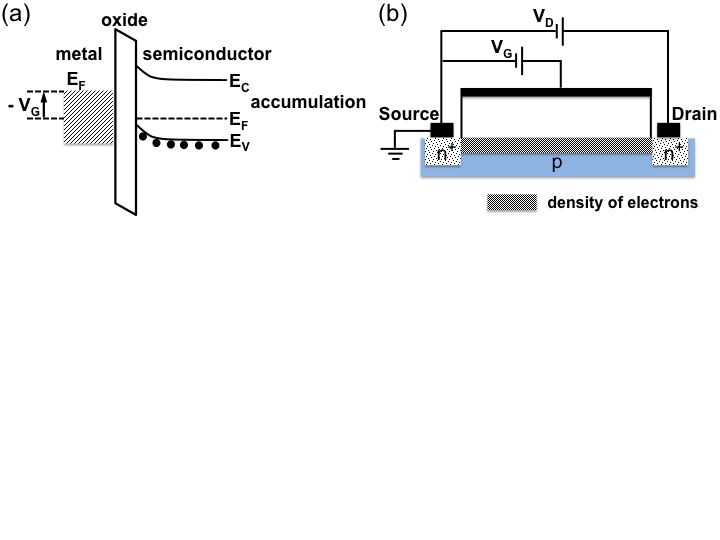}}
\vspace{-4.3truecm}
\caption{\footnotesize{%
(a) Metal-oxide-semiconductor (MOS) interface. An applied gate voltage, $V_G$,  changes the band bending and the carrier density at the MOS interface. 
The conduction electrons of the gate are depicted by the shaded region. (b) Schematic of a MOSFET with source and drain contacts made of heavily $n$-doped 
regions to ensure Ohmic contacts through a thin Schottky barrier.   
}}
\label{fig:FET}
\vspace{-0.2truecm}
\end{figure}

In the context of proximity effects, electrostatic gating has an important role by providing their tunability. 
The principle of such gating can be understood from field-effect transistors (FETs), central to conventional electronics. 
FETs rely on the electrostatic gating where the gate voltage controls the conductivity of the device. 
This electrostatic gating is illustrated in Fig.~\ref{fig:FET} on the example of metal-oxide-semiconductor (MOS) structure, 
which effectively acts as a capacitor,  and the MOSFET implementation. With TMDs it is possible to realize
atomically thin FETs of high on/off ratios~\cite{Radisavljevic2011:NN,Chuang2016:NL}.
The resulting gate-controlled carrier density can also profoundly transform materials properties turning an insulator into a superconductor~\cite{Ahn2006:RMP,Ueno2008:NM,Ye2012:S}. 
Even without changing the carrier density, the gate voltage could induce ferromagnetism in semiconductors~\cite{DiasCabral2011:PRB,Ahn2015:PRB}.  

\begin{figure*}
\resizebox{17.8cm}{!}{\includegraphics{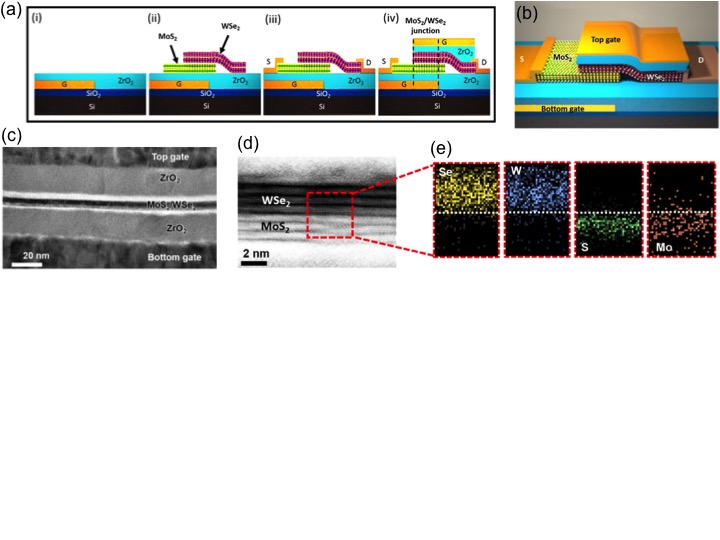}}
\vspace{-6.7truecm}
\caption{\footnotesize{%
(a) Schematic for the fabrication process  of a dual-gated MoS$_2$/WSe$_2$ diode. (i) Local bottom gate electrode with Ni as the metal electrode and ZrO$_2$ as the gate dielectric. (ii) MoS$_2$ and WSe$_2$ layers dry-transferred onto the bottom gate and etched to form a rectangular heterostructure. (iii) Metal contacts to MoS$_2$ (Ni) and WSe$_2$ (Pd) deposited, as source and drain electrodes, respectively. (iv) Top-gate stack with ZrO$_2$ as the gate dielectric and Ni metal as the electrode. Adapted with permission from Ref.~\cite{Roy2015:ACSN}.
}}
\label{fig:dual}
\vspace{-0.4truecm}
\end{figure*}

Given the 
short screening lengths, the influence of electrostatic gating is mostly an interface effect. This makes 2D systems, 
including vdW materials with atomically flat interfaces, 
suitable candidates for tuning their electronic structure by gating. 
However, one should also recognize the bonding character of a material to be modified by gating. With chemical 
bonding any tunability in the electronic structure and the density of states (DOS) is precluded~\cite{Lazic2016:PRB}. 
This situation is analogous to a superglue: 
the two bonded regions 
are strongly altered. Even though ion liquid gating can generate large fields $E_\mathrm{ext}  \sim 1$ V/\AA~\cite{Mannhart1993:APL,Hebard1987:IEEETM}, 
comparable to the strength of a chemical bond, breaking such a bond, similar to the superglue, leads to irreversible damage and eliminates tunability. 

In contrast, a much weaker vdW bonding  is  
analogous to the reversible character of the post-it note which can be attached and reattached to different locations.  A simple electrostatic model
for gating 2D systems with vdW bonding explains that an effective gating is a consequence of  a large  
dielectric constant which combines contributions of a small Femi level DOS 
and a large bonding distance~\cite{Lazic2016:PRB}.  An enhanced effective dielectric constant supports a gate-tunable electronic structure.  
For example, graphene's Dirac cone, can be reversibly moved by gating with respect to the Fermi level.

A convenient implementation for gating vdW materials is provided by dual-gate platform which enables an independent control of 
of the electrostatic potential and carrier density or, equivalently, the electric field and the position of the Fermi level~\cite{Roy2015:ACSN,Wang2017:NL}. 
A particular implementation of a dual-gate platform based on two semiconductor TMDs: MoS$_2$ and WSe$_2$ layers with high-quality and 
atomically sharp interfaces, is shown in Fig.~\ref{fig:dual}. Through changes of the electrostatic potential and carrier density a similar 
platform could enable tunable magnetic and spin-orbit proximity effects.   

\section{3. Magnetic Proximity Effects}\label{Sec:MPE}

\subsection{3.1 Spin Injection vs Magnetic Proximity}

Even though proximity effects usually imply equilibrium properties (zero applied bias), they can also 
alter the nonequilibrium behavior of materials.  
To better understand the distinction between equilibrium and nonequilibrium processes and the associated
lengthscales, we consider magnetic junctions, building blocks in the field of spintronics~\cite{Zutic2004:RMP,Fabian2007:APS} 
and the key elements
.in computer hard drives and magnetic random access memory (MRAM)~\cite{Tsymbal:2011}. 
The goal to manipulate spin degrees of freedom
often requires introducing spin-dependent properties in the material where they are initially absent, such that
spin up and spin down electrons (with respect to the direction of a magnetization or an applied magnetic field)
are no longer equivalent.

\begin{figure}[t]
\resizebox{8.7cm}{!}{\includegraphics{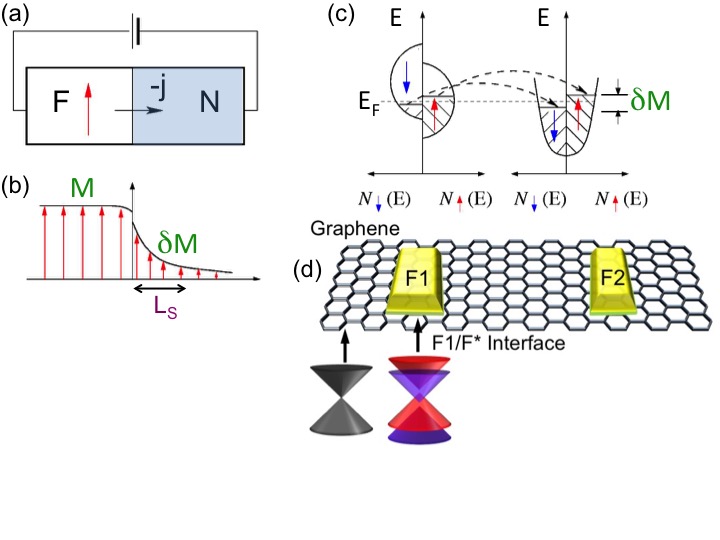}}
\vspace{-1.5truecm}
\caption{\footnotesize{%
(a) Schematic of spin injection from a ferromagnet (F) into a nonmagnetic region (N). Electrons
flow from F to N (opposite to the current j). (b) Spatial dependence of the magnetization $M$, nonequilibrium
magnetization $\delta M$  (spin accumulation) decays in N over the spin diffusion length, $L_S$. (c) Contribution of different spin-resolved DOS 
to both charge and spin transport across the F/N interface leads to $\delta M$. (d) Magnetic proximity effects 
in F1/ graphene junction. The electronic structure of proximity-modified graphene, F*, becomes spin-dependent. A ferromagnet, F2, 
could be used for detecting magnetic proximity effects through transport.}}
\label{fig:inject}
\vspace{-0.2truecm}
\end{figure}

Nonequilibrium spin is the result of some source of pumping arising from transport, optical, or resonance methods. Once the
pumping is turned off, the spin will return to its equilibrium value~\cite{Zutic2004:RMP}.
Electrical spin injection, a transport method for generating nonequilibrium spin, is shown in Fig.~\ref{fig:inject}(a)-(c).
A ferromagnet (F) has a net magnetization $M$ and inequivalent spin up and spin down DOS.
When a charge current flows across the F/nonmagnetic region (N) junction, spin-polarized carriers in a ferromagnet
contribute to the net current of magnetization entering N, resulting in the nonequilibirum magnetization $\delta M$, also
known as the spin accumulation.
A characteristic length scale for $\delta M$ is the spin diffusion length, $L_S$ $>$ 100 nm 
in many materials, 
while in graphene it  can even exceed 30 $\mu$m at 300 K~\cite{Drogeler2016:NL}.

\begin{figure}[t]
\resizebox{8.7cm}{!}{\includegraphics{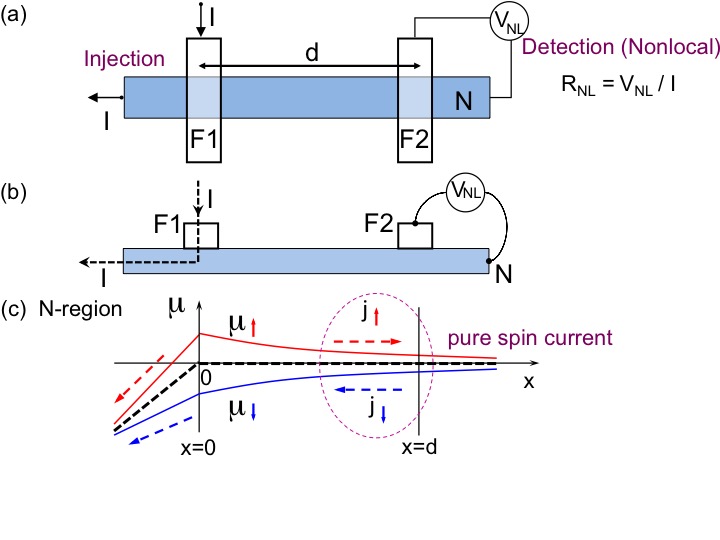}}
\vspace{-1.5truecm}
\caption{\footnotesize{%
Spin injection and nonlocal detection in a lateral spin-valve device. 
(a) Top and (b) side view. The bias 
current, $I$, flows from F1 to the left end of N, the spin signal is detected
by measuring a nonlocal voltage, V$_\mathrm{NL}$ between F2 and N. V$_\mathrm{NL}$ and the nonlocal 
resistance, R$_\mathrm{NL}$=V$_\mathrm{NL}$/I, depend on the relative orientation of {\bf M} in F1 and F2.
(c) A spatial dependence of electrochemical potential $\mu$ (broken line) and its spin-resolved components in N. For $x>0$,
there is no net charge current density, $j=j_\uparrow+j_\downarrow$, but as a result of spin diffusion and $\delta M$, 
only pure spin current, $j_\uparrow - j_\downarrow$, flows.
}}
\label{fig:nonlocal}
\vspace{-0.2truecm}
\end{figure}

Such  a spin accumulation and spin-polarized currents are readily detected by placing another F, i.e. in
the F1/N/F2 geometry.  The resulting approach is analogous to the polarizer-analyzer method of detecting the polarization of light 
propagating through two optical linear polarizers~\cite{Zutic2004:RMP}, shown in Fig.~\ref{fig:nonlocal}. Using a nonlocal geometry pioneered
by the work of Johnson and Slisbee~\cite{Johnson1985:PRL,Johnson1987:PRB}, spin injection is spatially separated 
from spin detection to eliminate spurious effects 
attributed to spin transport~\cite{Song2014:PRL,Maekawa:2012}. Driven by the spin accumulation and thus $\delta M$,  in the equipotential 
region $x >0$, there is a flow of pure spin current, $j_\uparrow-j_\downarrow$, with the spin-resolved current density,
$j_{\uparrow,\downarrow}$,  proportional to the slope of $\mu_{\uparrow,\downarrow}$, the spin-resolved electrochemical potential~\cite{Zutic2004:RMP}. 
The resulting spin-injection signal, $\delta M$, is detected by the nonlocal voltage or resistance in F2, spatially separated  from the injector F1. 
This approach of Johnson and Silsbee is frequently employed in vdW materials and
further discussed in experiments in Secs. 3.2, 3.3, 4.4, 5.3, and 5.4.

In contrast to $\delta M$, without any current flow (zero applied bias), there could still be some {\em equilibirum} magnetization in the N region 
through the magnetic proximity effect, but its typical lengthscale is less than nm~\cite{Hauser1969:PR}. Common understanding of the
spin injection in the F/N region implies that N is completely nonmagnetic with spin up and spin down DOS equivalent, a tiny
interface region where a magnetic proximity effect may be present is readily neglected in comparison with a much larger $L_S$. 

However, the situation is qualitatively different for an atomically thin N region.  The thickness of ML vdW materials is smaller than the characteristic magnetic proximity length and thus in such a geometry
 interface and proximity effects become crucial. 
A part of the N region next to the F is transformed by the magnetic proximity effects acquiring across its thickness 
equilibrium spin-dependent properties which also directly modify the nonequilibrium properties including the flow of current or
optical excitation in that region. The process of spin injection is no longer from the F to N region, but from F to the proximity modified region
F*.  For graphene, as shown in Fig.~\ref{fig:inject}(d), such a F* could lead to the proximity-induced exchange splitting of a Dirac cone. 
Consequently, the analysis of spin injection and nonlocal detection in Fig.~\ref{fig:nonlocal} could be strongly 
modified by proximity effects if N is an atomically thin region. The nonequilibrium (transport) properties, including the flow of 
charge and spin current, will depend on the proximity-induced exchange splitting in F* below F1~\cite{Lazic2014:PRB,Lee2016:PRB}.

It is helpful to distinguish two mechanisms for magnetic proximity effects~\cite{Lazic2016:PRB}: (i) The wave
functions from graphene penetrate into the insulating F 
as evanescent states since there are no states there at the Fermi level, 
where they acquire exchange splitting from its native ferromagnetism.
(ii) The wave functions from the metallic F penetrate into graphene,
directly polarizing its electronic structure at the $E_F$.

At the time of an early work on magnetic proximity effects, there was a 
considerable interest to study the influence of a magnetic impurity in metals~\cite{Coey:2009}
The outcome, similar to magnetic proximity effects, is material-specific
and depends on the local environment. The same magnetic impurity placed
in a different nonmagnetic matrix can lead to very different results.  Co  
placed in Al loses its magnetic moment, retains it in Cu, while in Pd it can even
lead to the formation of a giant moment, tens of Bohr magnetons. A reduced
magnetic moment is also associated with the screening in the Kondo effect~\cite{Coey:2009, Khomskii:2010}.

\begin{figure*}
\resizebox{18cm}{!}{\includegraphics{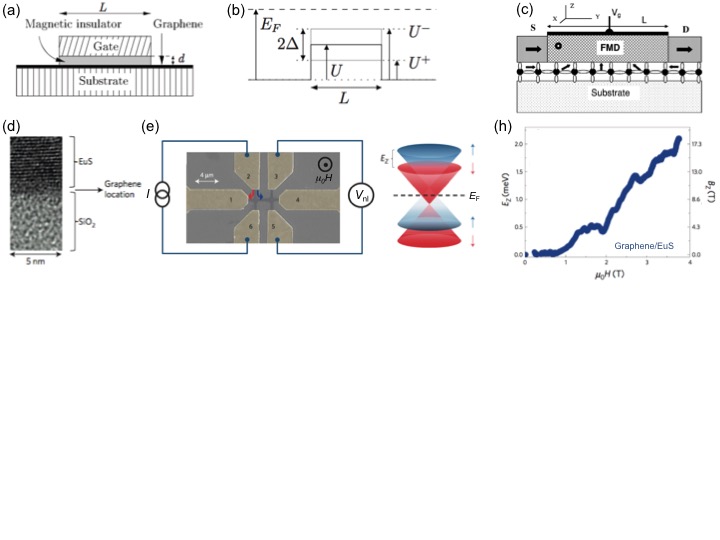}}
\vspace{-7.1truecm}
\caption{\footnotesize{%
(a) F insulator 
induces an exchange splitting, $\Delta$, in graphene. A metallic gate controls the electrostatic potential.
(b) Ferromagnetic proximity effect splits the barrier: $U^\pm = U \mp \Delta$~\cite{Haugen2008:PRB}. 
(c) Schematic of the spin FET utilizing a graphene channel  (circles with bonds)  and a ferromagnetic dielectric (FMD). The source  S  and drain  D  have collinear 
magnetizations, perpendicular to the one in FMD. The electron spin (small arrow)  precession
due to the exchange interaction with FMD. $V_g$  alters the exchange interaction and
the resulting precession rate~\cite{Semenov2007:APL}.
(d) Left: A TEM cross-sectional image of a graphene/EuS  showing a sharp interface.
(e) A SEM device image. The central Hall bar region:  graphene coated with EuS. The outer regions (1-6): Ti/Pd/Au electrodes. Non-local measurements are carried out by applying current $I$ along leads 2 and 6 and measuring non-local voltage $V_\mathrm{nl}$ between
leads 3 and 5. The applied field  $\mu_0$H directs the oppositely spin-polarized charge carriers towards opposite directions along the Hall bar channel, 
spin-up (spin-down) current: the blue (red) arrows. Right: Zeeman splitting of the Dirac cone and the Fermi level, $E_F$.  
(f) Quantitative estimation of the Zeeman splitting energy $E_Z$. On top of the main curve, secondary structures 
may be attributed to the multi-domain magnetization process of EuS. The right axis: the estimated total Zeeman field ($B_Z$) 
in graphene enhanced by the EuS-induced interfacial exchange field~\cite{Wei2016:NM}. Adapted with permission (a), (b) from Ref.~\cite{Haugen2008:PRB}, (c) from Ref.~\cite{Semenov2007:APL}, and
(d)-(f) from Ref.~\cite{Wei2016:NM}}}.
\label{fig:GFI}
\vspace{-0.2truecm}
\end{figure*}

\subsection{3.2. Proximity with Magnetic Insulators}

Functionalizing graphene by adatoms or vacancies, shown in Fig.~1 provides an example of how magnetism and spin-dependent properties 
can be introduced in various 2D materials~\cite{Han2014:NN,Yazyev2007:PRB,McCreary2012:PRL}. 
In contrast to this local and random creation of magnetic moments, placing graphene 
on a magnetic substrate provides a very different approach by realizing controllable and a more uniform proximity-induced magnetism. 
The choice of magnetic insulators, such as EuS, EuO, EuSe, or yttrium-iron garnet (YIG), appears particularly suitable for magnetic proximity effects in 2D materials. 
Eu-based compounds have been extensively studied including the first demonstration of a solid-state spin-filter~\cite{Zutic2004:RMP,Esaki1967:PRL}, 
giant spin-splitting~\cite{Tsymbal:2011}, and the spin-dependent tunneling current in the F/superconductor junctions~\cite{Moodera1988:PRB,Xiong2011:PRL,Li2013:PRL} 
while YIG with it high $T_C\sim 550$ K is a widely used ferrimagnet~\cite{Jiang2015:NL,Jungfleisch2016:PRL}.

With only expected weak hybridization, largely preserving the native electronic structure of the nearby 2D materials, these insulating ferro/ferrimagnets have motivated 
several theoretical proposals for using magnetic proximity effects~\cite{Haugen2008:PRB,Semenov2007:APL,Lazic2016:PRB,Scharf2016:PRL}, 
including those illustrated in Fig.~\ref{fig:GFI}. 
F insulators, such as EuO, could induce gate-controlled exchange splitting, $\Delta$, 
in the neighboring graphene layer and modify its transport properties. The resulting spin-dependent barrier formed by the F insulator shifts differently the bottom of the conduction 
band for spin up/down, as shown in Figs.~8(e) and (h). 
As a consequence, the total conductance across the barrier will be spin-polarized which could be detected by 
measuring magnetoresistance (MR) in a spin-valve geometry with an added F region, as in Fig.~7.  Nominally, there is similarity with such proximity-induced 
exchange splitting and a Zeeman splitting  
$\approx 2\mu_B B$,  from an applied in-plane magnetic field, $B$. However, small $g$-factors in graphene require huge applied fields: 20 T would only yield spin-splitting of  $\approx 1$ meV.

In a variation of a spin FET proposal by Datta and Das~\cite{Zutic2004:RMP,Datta1990:APL} (discussed further in Sec.~5.3), 
the spin rotation of the spin-polarized carriers traveling between the F 
source and drain, 
would be controlled by proximity-induced exchange interaction in the graphene nanoribbon channel.  Perpendicular magnetization in the F insulator (gate dielectric) 
with respect to the collinear direction of {\bf M} in the source and drain sets the precession of the carrier spin with the rate controlled by the gate voltage. The outcome of this scheme would be a gate-controlled 
source-drain conductance, determined by the alignment between the carrier spin entering drain and its 
{\bf M}~\cite{Semenov2007:APL}. 

Despite the conceptual simplicity of using F
insulators, they present considerable materials challenges. EuO, often preferred to EuS due to its higher ferromagnetic Curie temperature (69 K vs 16 K in the bulk), required complex synthesis to be first integrated with graphene~\cite{Swartz2012:ACSN} EuO is not thermodynamically stable and easily converts to nonmagnetic Eu$_2$O$_3$. 5 nm EuO, capped with 2 nm MgO, grown on graphene has revealed epitaxial growth with (001) orientation and a large Kerr angle, consistent with a magneto-optic response of high-quality EuO thin films. Eu-based magnetic insulators are also challenging to describe theoretically for proximity effects. Their simple interfaces with 2D materials are polar and undergo surface reconstruction thus altering the values of their exchange interaction parameters that would be deduced from commonly employed models~\cite{Qi2015:PRB}.

The magnitude of the proximity-induced exchange field, $B_\mathrm{ex}$, strongly depends on the quality of the interface and F insulator. 
A high-quality graphene/EuS heterostructure in Fig.~\ref{fig:GFI}(d)-(f) demonstrates a strong proximity-induced modification of transport properties~\cite{Wei2016:NM}. 
The exchange splitting of the Dirac cone generates electron- and hole-like carriers at the Dirac point. Under a Lorentz force these carriers propagate 
in opposite directions and yield a pure spin current and a non-local voltage, $V_\mathrm{NL}$. In spin transport it is convenient to study the nonlocal resistance (recall Fig.~7), 
$R_\mathrm{NL} \equiv V_\mathrm{NL}/I$~\cite{Johnson1985:PRL,Zutic2004:RMP}, $R_\mathrm{NL}=R_0+ \beta(\mu_0 H) E_Z^2$ is evaluated at its peak value at the Dirac point,  
$\beta$ represents the orbital-field effect, while
the Zeeman splitting energy, $E_Z$, is expressed in terms of the total Zeeman field, $B_Z$, 
$E_Z = g \mu_B B_Z=g \mu_B (B_\mathrm{ex} +\mu_0 H)$, dominated by the exchange contribution, $B_\mathrm{ex} > 14$ T, as estimated 
from Fig.~\ref{fig:GFI}(f)~\cite{Wei2016:NM}. This large $B_\mathrm{ex}$ also lifts the ground-state degeneracy of graphene in the quantum Hall regime which is reached
at $\mu_0 H \sim 3.8$ T, confirming that exotic materials properties can be realized at much smaller applied magnetic fields than what is required
without magnetic proximity effects. For example, in high-quality graphene, the quantum Hall effect was observed for 
an in-plane field  of $\mu_0 H > 20$ T~\cite{Wei2016:NM}. 

\begin{figure*}[!t]
\resizebox{18cm}{!}{\includegraphics{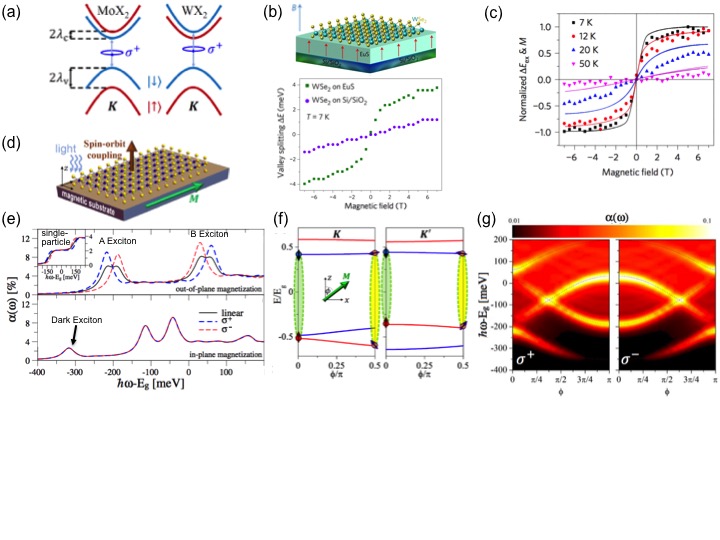}}
\vspace{-4.1truecm}
\caption{\footnotesize{%
(a) Spin-valley coupling. CB and VB are spin split in the K valley by the SOC $2\lambda_{c,v}$.
The emitted/absorbed light have valley-selective helicity $\sigma^\pm$.
(b) ML WSe$_2$ 
on magnetic substrate EuS shows the valley splitting in the perpendicular magnetic field strongly enhanced by  
proximity effect. 
Similar results were observed with CrI$_3$~\cite{Zhong2017:SA}. 
(c) Field-dependent valley-exchange splitting, $\Delta E_\mathrm{ex}$ of WSe$_2$ only due to magnetic proximity effect and $M$ of EuS, normalized to their saturated values at 7 K.  
Their mutual agreement within the F state of EuS at  7 and 12 K, confirms
that 
the enhanced valley splitting is caused by the 
magnetic proximity effect. (d) ML TMD on a magnetic substrate. 
(e) Absorption spectra of MoTe$_2$ on EuO for different polarizations with out-of-plane and in-plane exchange splitting. 
The inset: single-particle absorption. $\lambda_c=-18$ meV,  $\lambda_v=110$ meV, exchange splitting $J_c =100$ meV and $J_v=85$ meV.
(f) The $K$ and $K'$ band edges as {\bf M} is rotated, shown for MoTe$_2$/EuO parameters. One dark exciton for $K$ and $K'$ and the spin direction 
for selected band edges are depicted.
(g) Evolution of the absorption 
as {\bf M}  is rotated from out of plane ($\phi=0$) to in plane ($\phi=\pi/2$) and out of plane, but with reversed {\bf M} ($\phi=0$), 
parameters as in 
(e). Adapted with permission (a), (d)-(g) from Ref.~\cite{Scharf2017:PRL}, (b) and (c) from Ref.~\cite{Zhao2017:NN}.
}}
\label{fig:TMD}
\vspace{-0.2truecm} 
\end{figure*}

Strong magnetic proximity effects, up to $\sim 300$ K,  have been observed in graphene on YIG by measuring anomalous Hall effect,
consistent with the proximity-induced {\bf M} in graphene~\cite{Wang2015:PRL}. With a high-quality YIG interface, the mobility of graphene 
was comparable or even higher than in graphene/SiO$_2$ devices~\cite{Wang2015:PRL}. This undiminished mobility was in contrast to  
using doping or functionalization to introduce {\bf M} in a nonmagnetic region.

In a bilayer graphene on YIG it was demonstrated that by  changing the in-plane direction of its {\bf M}  the spin current in a lateral
spin valve device can be strongly modulated~\cite{Singh2017:PRL}. From the strong temperature dependence of the nonlocal spin signal 
an additional contribution to spin relaxation in graphene could be attributed to thermally induced transverse fluctuations of {\bf M} in YIG
as well as estimate the lower bound of the proximity-induced magnetic exchange field to be approximately 1 T~\cite{Singh2017:PRL}.  
It was predicted that a similar change of the in-plane {\bf M}  in YIG, together with the strong SOC in a nearby topological insulator,  
could yield novel Hall effect with a maximum transverse voltage when the current is parallel to {\bf M} and the previous Hall effecs
were expected to vanish~\cite{Scharf2016:PRL}.

In addition to changes in transport, proximity effects can also strongly alter optical properties in many materials. This is particularly pronounced in ML TMDs, 
MX$_2$ (M =Mo, W, X = S, Se, Te),  which have unique optical properties that combine a direct band gap, very large binding energies (up to $\sim$0.5 eV) 
for excitons (bound electron-hole pairs), and efficient light emission~\cite{Mak2010:PRL,Chernikov2014:PRL}. 
Unlike graphene, TMDs have a large band gap and a strong SOC 
due to the $d$ orbitals of the heavy metal atoms and broken inversion symmetry. One of their hallmarks is the strong valley-spin coupling~\cite{Xiao2012:PRL} which 
leads to a valley-dependent helicity [opposite for $K$ and $K'$ valley, see Fig.~3(c)] 
of optical transitions between conduction and valence band (CB, VB), shown 
in Fig.~\ref{fig:TMD}(a) with a reversed CB ordering for MoX$_2$ and WX$_2$. 
The SOC Hamiltonian can be written as $H_\mathrm{SO}=\bm{\Omega}(\bm{k})\cdot\bm{s}$ using the SOC field $\bm{\Omega}(\bm{k})$~\cite{Zutic2004:RMP,Fabian2007:APS}, 
where $\bm{k}$ is the wavevector and $\bm{s}$ is the vector of spin Pauli matrices. In ML TMDs, this leads to $\bm{\Omega}(\bm{k})=\lambda(\bm{k})\hat{\bm{z}}$, 
where $\lambda(\bm{k})$ is odd in $\bm{k}$ and $\hat{\bm{z}}$ is the unit vector normal to the ML plane. At the $K$ point, $\lambda(\bm{k})$ reduces to the values 
$\lambda_\mathrm{c(v)}$  CB (VB),  $\lambda_c$ is often neglected. Conventions also differ what is considered as $K$ and what as $K'$ valley~\cite{Xiao2012:PRL,Mak2016:NP}. 

Similar to lifting the spin degeneracy in spintronics, lifting the $K$/$K'$ valley degeneracy is crucial in manipulating valley degrees of freedom. 
A small Zeeman splitting of $\sim$0.1$-$0.2 meV/T in TMDs~\cite{Srivastava2015:NP,MacNeill2015:PRL,Stier2016:NC} and very large applied magnetic fields required for the removal of this
degeneracy can be overcome by magnetic proximity effects. Experiments using EuS magnetic substrate, 
shown in Fig.~\ref{fig:TMD}(b) and (c),
confirm a significant valley splitting which is manifested as the circularly polarized photoluminescence or reflectance spectra, dominated by 
excitons~\cite{Mak2010:PRL,Chernikov2014:PRL,Xiao2012:PRL,Mak2016:NP}. 
A small perpendicular applied field was needed to rotate 
 {\bf M}  out-of-plane for 
allowed optically transitions, equivalent to what is know for quantum wells~\cite{Zutic2004:RMP}.
The resulting valley splitting exceeding 2 meV at 1 T in WSe$_2$ using EuS~\cite{Zhao2017:NN}
and CrI$_3$~\cite{Zhong2017:SA} substrates 
is an order of magnitude larger than what would be possible with just an applied field, as well as much larger
than the spin splitting in graphene from Fig.~8(f). In fact, unpublished results show even a much 
larger proximity-induced valley splitting of $\sim$ 20 meV at 1 T for WS$_2$/EuS~\cite{Petrou2017:P}.

Until recently~\cite{Scharf2017:PRL}, 
magnetic proximity effects in a wide class of materials were only studied within the single-particle picture, 
neglecting the Coulomb interaction. Guided by this 
picture~\cite{Qi2015:PRB}, experimental efforts in TMD/F heterostructures have focused on the out-of-plane {\bf M}
which removes the valley degeneracy~\cite{Zhao2017:NN,Zhong2017:SA}. While excluding Coulomb interaction prevents calculating 
the position and spectral weight of excitons that were used to study magnetic proximity effects, some trends could be understood. 
In Fig.~\ref{fig:TMD}(a) 
there are optically allowed (forbidden) dipole transitions with a parallel (antiparallel) electron spin configuration, 
known as the bright (dark) excitons. The occurrence of lower- and higher-energy bright (A and B) excitons, 
schematically corresponding to the transition between blue (marked) and red CB and VB sub-bands, respectively. 

For an out-of-plane {\bf M}  which is collinear with $\bm{\Omega}(\bm{k})$ [see Fig.~\ref{fig:TMD}(d)], this simple picture of A and B excitons can be generalized 
expecting their proximity-induced exchange splitting in the helicity-resolved spectral response.
This experimentally observed behavior with 
the opposite proximity induced shift for A and B excitons~\cite{Zhao2017:NN,Zhong2017:SA}
is confirmed by an accurate inclusion of the Coulomb interaction for reflectance spectra from Fig.~\ref{fig:TMD}(e)~\cite{Zhao2017:NN}. 
In contrast, the 
single-particle 
picture [the inset of Fig.~\ref{fig:TMD}(e)] fails to capture any signs of excitons. 

In the seemingly trivial case of an in-plane  {\bf M}, where a single-particle description implies no lifting of the
valley degeneracy~\cite{Qi2015:PRB}, calculated absorption spectra show that dark excitons can become bright. This conversion between dark and bright excitons can be 
understood from the rotation of {\bf M}, generally non-collinear with SOC field, showing the evolution of the spin directions of the carriers forming the 
dark excitons and the corresponding absorption spectra for $K$ and $K'$ valleys. While the parameters were chosen for the MoTe$_2$/EuO heterostructure
with large CB and VB exchange splitting, the occurrence of dark excitons for in-plane {\bf M} is robust. It persist even for much 
smaller exchange splitting and is largely unchanged by the value of the interfacial SOC that can vary for different TMD/F junctions~\cite{Scharf2017:PRL}.

Recent advances in vdW materials show that even a single atomic layer can support 2D ferromagnetism
in insulating (CrGeTe$_3$, CrI$_3$, CrSiTe$_3$,...)~\cite{Gong2017:N,Huang2017:N,Lin2015:JMCC} 
and metallic (Fe$_3$GeTe$_2$, VSe$_2$,...)~\cite{Liu2017:NPJ2DMA,Bonilla2018:NN} forms. 
Precluding such 2D ferromagnetism based on the Mermin-Wagner theorem~\cite{Mermin1966:PRL} is not 
relevant as it assumes an isotropic magnetic system and the absence of spin-orbit coupling. In contrast, these layered
systems typically display a strong magnetocrystalline anisotropy~\cite{Zhuang2015:PRB}. 
 A list of additional ML vdW magnets and their potential use is given in Ref.~\cite{Duong2017:ACSN}. With the improvement in the growth
techniques~\cite{Zhou2018:N} and the understanding which of the materials can be exfoliated as MLs~\cite{Mounet2018:NN}
the number of available 2D vdW ferromagnets keeps increasing.  

While CrI$_3$ used for magnetic proximity effects~\cite{Zhong2017:SA} 
was $\sim$ 10 nm thick and not in the ML limit, an obvious next step would be to consider proximity effects with ML vdW ferromagnets. This approach is further 
supported by a very large low-temperature tunneling magnetoresistance (TMR) observed in ML 
CrI$_3$-based vdW heterostrucutres~\cite{Kim2018:P,Klein2018:P,Song2018:P,Wang2018:P} in which there is also
a prediction of a strong proximity-induced magnetization~\cite{Zhang2018:PRB}.

So far, magnetic proximity effects in TMDs employing F insulators were measured at cryogenic temperatures. However, this is not a fundamental limitation: 
Common F metals could enable room temperature proximity effects, while the metal/ML TMD hybridization can be prevented 
by inserting a thin insulating layer.  Unlike magnetic fields of $\sim$ 30 T~\cite{Molas2017:2DM,Zhang2017:NN} that exceed 
typical experimental capabilities, the removal of valley degeneracy using magnetic substrates is not complicated 
by orbital effects and yet could enable even larger valley splittings~\cite{Qi2015:PRB}. 
Magnetic proximity offers another way to control and study many-body 
interactions in the time-reversed valleys of ML TMDs. For example, by competing with the influence of the intrinsic SOC, it would change the energy of 
shortwave plasmons~\cite{VanTuan2017:PRX} put forth as an explanation for the low-energy dynamic band observed in W-based electron-doped TMDs~\cite{VanTuan2017:PRX,Jones2013:NN}.

Proximity to a ferro- or ferrimagnetic insulator essentially turns a normal metal into a ferromagnet, enabling the ``anomalous'' transport effects that become possible if the time-reversal symmetry is broken, such as the anomalous Hall and Nernst effects observed in a Pt film on YIG~\cite{Huang2012:PRL}. Metals like platinum and palladium are the most suitable for observation of such effects, because they are close to the Stoner instability and therefore have a large magnetic susceptibility.

\subsection{3.3. Proximity with Magnetic Metals} 

Studies of superconducting proximity effects (see Fig.~2) have both guided a common understanding of magnetic proximity effects 
and been used to provide early measurements of the characteristic length over which spin polarization from a F metal would penetrate
into a nonmagnetic region~\cite{Hauser1969:PR}. With a large magnetic susceptibility of a metal Pd, that length penetrating from 
Fe and Cr was enhanced to  $\sim$ 2 nm. The presence of such proximity-induced spin polarization could influence the second F
and was important in the early development of giant magnetoresistive devices~\cite{Zutic2004:RMP}.  

\begin{figure}[ht]
\vspace{0.08cm}
\resizebox{8.7cm}{!}{\includegraphics{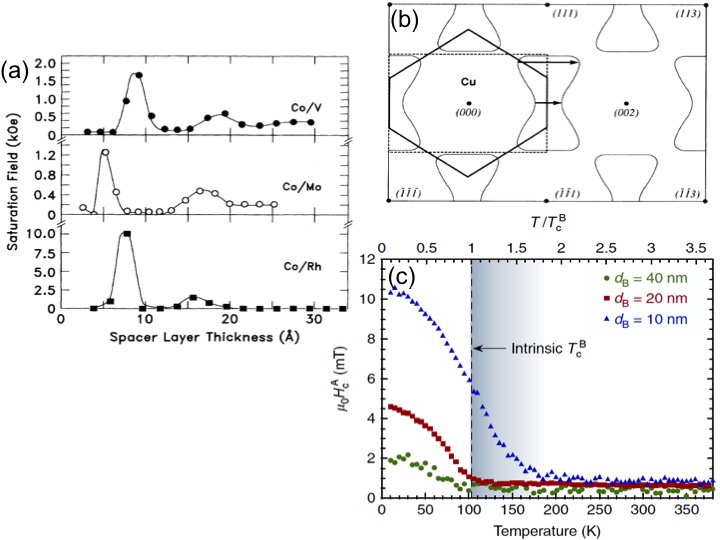}}
\vspace{-0.55cm}
\caption{\footnotesize{%
(a) Oscillation of the saturation field as a function of spacer-layer thickness in Co/V, Co/Mo, and Co/Rh multilayers~\cite{Parkin1990:PRL}.
(b) Fermi surface spanning vectors for FCC Cu with the (111) orientation of the interface, which determine the oscillation periods.
(c) Temperature dependence of the coercive field in the Co$_{85}$(AlZr)$_{15}$/Co$_{60}$(AlZr)$_{40}$/Sm$_{10}$Co$_{90}$ trilayer; 
the dashed line shows the Curie temperature of the middle layer. IEC between the top and bottom layer is seen well above that 
Curie temperature for the 10 nm thickness of the middle layer.
Adapted with permission (a) from Ref.~\cite{Parkin1990:PRL}, (b) from Ref.~\cite{Bruno1991:PRL}, (c) from Ref.~\cite{Magnus2016:NC}.
 }}
\label{fig:IEC}
\vspace{-0.2truecm}
\end{figure}

A thin layer of a normal metal separating two F layers mediates the so-called interlayer exchange coupling (IEC)~\cite{Grunberg1986:PRL,Majkrzak1986:PRL,Salamon1986:PRL}. 
This coupling can be simply viewed as a manifestation of the magnetic proximity effect induced by both Fs inside the normal metal. However, the physical picture is more complicated, 
because the perturbations induced in a metal have an oscillatory character, similar to Friedel oscillations, with a period that varies between different Bloch states. 
Such oscillations of the proximity-induced spin polarization  were already recognized over 50 years ago
in degenerate 2D electron gas and related to the Rutherman-Kittel-Kasuya-Yoshida (RKKY) 
interaction between magnetic impurities~\cite{Bardasis1965:PRL}.
As a result, the IEC oscillates as a function of the thickness of the normal layer~\cite{Parkin1990:PRL,Unguris1994:JAP}, shown in Fig.~\ref{fig:IEC},
and the periods of these oscillations correspond to the critical spanning vectors of the normal metal's Fermi surface~\cite{Bruno1991:PRL,Stiles:2005},
while the ferromagnetic ordering temperature can be changed through such coupling~\cite{Bovensiepen1998:PRL}.

\begin{figure}[ht]
\resizebox{14cm}{!}{\includegraphics{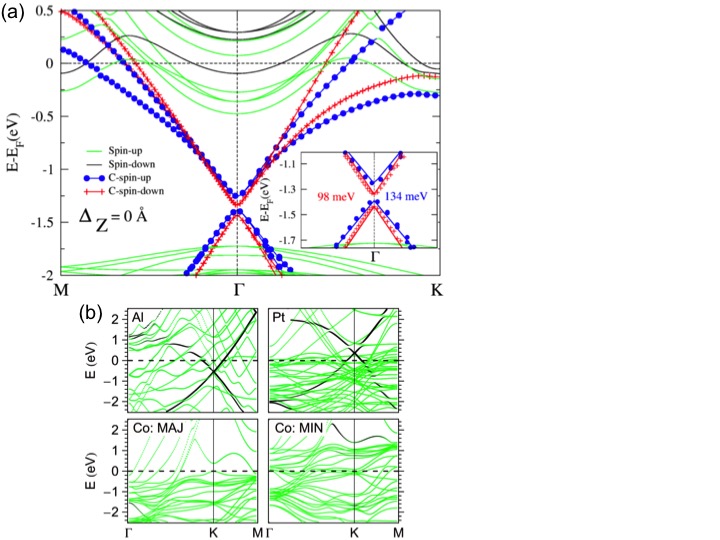}}
\vspace{-0.4cm}
\caption{\footnotesize{%
(a) Band structure of graphene on EuO. Green (blue) and black (red): spin up and spin down bands of EuO (graphene). 
Inset: zoom around the Dirac cone, the symbols: DFT data, the lines: dispersion fit~\cite{Yang2013:PRL}.
(b) Band structure of graphene on Al, Pt, and Co (111) substrates~\cite{Giovannetti2008:PRL}. The bottom left (right) panel corresponds to majority  (minority) spin band structure. 
The Fermi level is at zero energy. The amount of carbon $p_z$ character is indicated by the blackness of the bands. The conical point corresponds to the crossing of predominantly $p_z$ 
bands at K~\cite{Giovannetti2008:PRL}. Adapted with permission (a) from Ref.~\cite{Yang2013:PRL} and (b) from Ref.~\cite{Giovannetti2008:PRL}.
}}
\label{fig:MM}
\vspace{-0.2truecm}
\end{figure}

\begin{figure*}
\resizebox{18cm}{!}{\includegraphics{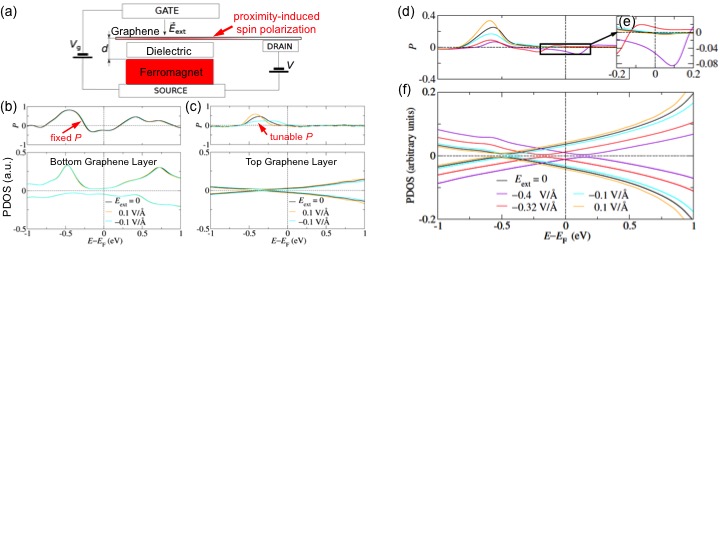}}
\vspace{-7.8truecm}
\caption{\footnotesize{%
(a) Schematic of the lateral device geometry. The red color depicts a proximity-effect induced DOS spin polarization, $P$, 
in graphene. (b), (c) Field-dependent $P$ and projected DOS (PDOS) on the C atoms, for the bottom 
(chemically bonded to the Co surface) and top graphene layer (vdW bonded to the bottom Gr layer).
(d) $P$  on the C atoms of Co/h-BN/graphene heterostructure. (e) Zoom of $P$ from (d) near $E_F$. 
(f) Field-dependent PDOS on the C atoms. Adapted with permission from Ref.~\cite{Lazic2016:PRB} 
}}
\label{fig:FGate}
\vspace{-0.2truecm}
\end{figure*}

Despite diminishing research on IEC, recent work on extending the range of magnetic proximity effects up to $10$ nm 
with a paramagnet separating F1, F2~\cite{Magnus2016:NC},  and  the intriguing possibility to switch  magnetization without 
an applied magnetic field~\cite{Lau2016:NN}, 
highlight the opportunities in transferring ideas of magnetic proximity effects with F metals to other materials systems.

A robust room temperature ferromagnetism in metals 
Co, Fe, and Ni, could be.valuable for proximity effects with vdW materials. 
However, direct metal contacts with graphene pose important challenges which could be understood in comparison with F insulators in Fig.~\ref{fig:MM}.
For an idealized graphene/EuO heterostructure, neglecting the reconstruction of the polar interface, first-principles calculations suggest that the linear 
band structure of the graphene is largely preserved~\cite{Yang2013:PRL}. The Dirac cone is clearly visible, but it acquires a spin-dependent gap, $\Delta_\sigma$, and Fermi velocity, 
$v_\sigma$, which can be fitted close to the Dirac point by the dispersion, $E_\sigma(q)=\pm[(\hbar v_\sigma q)^2+(\Delta_\sigma/2)^2]^{1/2}$, 
where $\Delta_\uparrow=134$ meV, $\Delta_\downarrow=98$ meV, $v_\uparrow=1.15 \times 10^6$ m/s, and $v_\downarrow=1.4 \times 10^6$ m/s~\cite{Yang2013:PRL}.
A relatively weak graphene/F interaction and hybridization is also expected from a calculated equilibrium distance of 2.57 $\mathrm{\AA}$  between graphene 
and EuO, considerably larger than $\approx$ 2.1 $\mathrm{\AA}$ if Ni or Co was used instead.

For metallic contacts two cases can been seen in Fig.~\ref{fig:MM}(b)~\cite{Giovannetti2008:PRL}. A weaker bonding with Al or Pt still preserves the Dirac cone, albeit largely shifted
below or above from the Fermi level, $\sim 0.5$ eV (depending on the relative difference between their work function with respect to the one for graphene),
signaling doped graphene. The key properties of graphene associated with the Dirac cone become largely inaccessible since the heterostructure will be dominated
by the electronic structure close to the Fermi level. The case of graphene on Co reveals a much stronger hybridization, very similar to the graphene on Ni(111)~\cite{Lazic2014:PRB}.  
The Dirac cone is completely destroyed, as expected  for a typical example of chemical bonding leading to a new interfacial material, distinct from the constituents 
in the original heterostructure.

An additional challenge~\cite{Yang2013:PRL} of using F metals as substrates is that they short-circuit the graphene layer 
and limit the design of possible devices.  While one could raise the same concerns for using ferromagnets with other 2D materials in which 
proximity effects would be induced, this is not a fundamental obstacle. Simply inserting a dielectric between the F metal and graphene or 
another 2D material could overcome these perceived difficulties. This approach~\cite{Lazic2016:PRB},
depicted in Fig.~\ref{fig:FGate}(a), offers several important opportunities. (i) Such systems include vdW heterostructures with atomically sharp 
interfaces [16], which simplify the implementation of electrostatic gating [2,17]; and (ii) these are key building blocks for graphene spintronics [18] 
with a prospect of gate-tunable magnetic proximity effects.

Bilayer graphene on F metal is a suitable system to consider the viability of gating by examining the influence of electric field on the layer-resolved DOS. 
In contrast to negligible DOS changes of the bottom graphene layer, the changes with gating in the top layer 
are considerable, as shown Fig.~\ref{fig:FGate}(b) and (c). This confirms a trend that strongly bonded heterostructures are unsuitable for gating: The chemical bonds 
ground the attached dielectric to the metallic F [Fig.~\ref{fig:FGate}(a)], precluding charge transfer and control of DOS spin polarization,
$P=(N_\uparrow-N_\downarrow)/(N_\uparrow+N_\downarrow)$.

Intuitively, a large bonding distance could provide a large voltage drop, while small DOS suppresses screening of the external field $E_\mathrm{ext}$. The resulting 
charge transfer for the region (top graphene layer) with a small DOS at the Fermi level $N(E_F)$ will induce appreciable changes in its electronic structure. 
Thus, to facilitate the tunability of $P$, one should seek an energy window with a small DOS in both spin channels. In Fig.~\ref{fig:FGate}(c) this is observed at 
 $\sim0.4$ eV below the $E_F$ for the vdW-bound top graphene layer, where the Dirac cone is largely preserved.

To predict the gating effects in systems similar to the one shown in Fig.~\ref{fig:FGate}(a), but not limited to graphene as the top layer, a simple electrostatic model
yields an estimate the DOS  shift for graphene relative to the F (the ``ground") when  $E_\mathrm{ext}$ is applied by the gate. With bottom (top) layers denoted
by $1$ and $2$, in the relevant limit of $N_1 \gg N_2$ (with chemical bonding, the bottom layer ``1"  becomes a part of the F metal), the relative shift of 
the electrostatic potential under gating, $\delta V = \delta  V_2 - \delta V_1$, with $\delta V= \epsilon_0  E_\mathrm{ext} d/(\epsilon + e^2 N_2 d)$, shows
that a small $N_2$ is required to achieve effective gating, while a large bonding distance, $d$ [see Fig.~\ref{fig:FGate}(a)], is desirable.
In a simple picture, chemical bonding (chemisortption) can be viewed as  a superglue, preventing gate-controlled changes in the DOS structure 
until an extremely large $E_\mathrm{ext}$ breaks the bond. In contrast, vdW bonding (physisorption) behaves as a post-it note which
can be moved and re-attached to another position, depicting the gate-tunable changes in the electronic structure.   These trends in gating
and suitability of vdW bonding have been corroborated not only by considering other layered systems, such as benzene that unlike graphene has a 
nonperiodic structure, but also for single atoms, such as Xe~\cite{Lazic2016:PRB}.

\begin{figure*}
\resizebox{17.8cm}{!}{\includegraphics{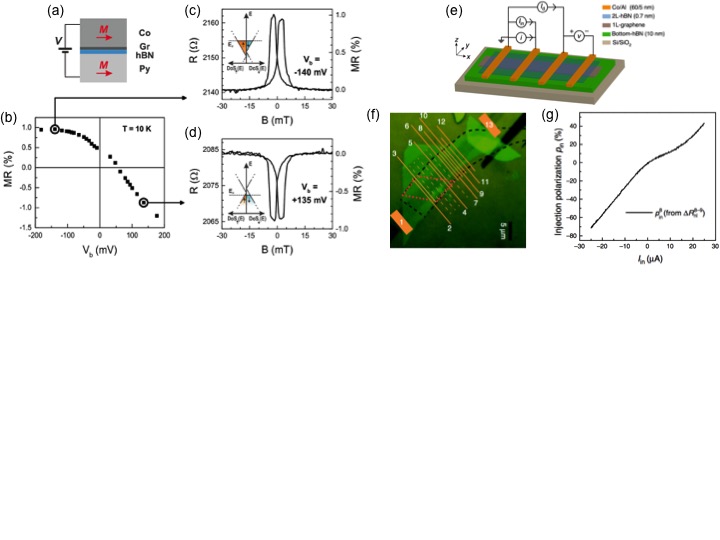}}
\vspace{-7cm}
\caption{\footnotesize{%
(a) Schematic for Co/G/h-BN/Py device showing a sign reversal of MR under bias, $V_b$. (b) MR as a function of $V_b$, for Co/G/h-BN/Py with Co and 
Py separated by a ML graphene and a bilayer h-BN. (c) MR for maximum positive $V_b$ and (d) maximum negative $V_b$. The MR 
sign reversal 
at $V_b\approx +60$ meV 
yields a shift in graphene's Fermi level from the conduction to valence band~\cite{Asshoff2017:2DM}.  
(e) Schematic of the vdW heterostructure of the 2L-h-BN/graphene/thick-h-BN stack with F Co electrodes. Nonlocal spin transport 
scheme with a DC current $I_\mathrm{in}$
and AC current, $i$, applied across the injector and a non-local differential (AC) spin signal $v$. 
(f) An optical microscopic picture of the vdW heterostructure. The black-dashed (red-dashed) line outlines the h-BN tunnel barrier.
The deposited Co electrodes (orange bars) and the Co/h-BN/graphene contacts are denoted by 1,...,13. The orange-dashed lines: 
unused contacts. Co electrodes 2-5 are 
fully or partially deposited on top of the ML region of the tunnel-barrier, 
the electrodes 6-12 are 
deposited on the bilayer region. The width of the Co electrodes (2-12) is between 0.15 and 0.4 $\mu$m.
(g) Differential spin-injection polarization of the injector contact 8, $p^8_\mathrm{in}$ as a function of $I_\mathrm{in}$,
calculated from the $\Delta R_\mathrm{nl}^{8-9}$$(I_\mathrm{in})$. Adapted with permission (a)-(d) from Ref.~\cite{Asshoff2017:2DM}, (e)-(g) from Ref.~\cite{Gurram2017:NC}.}}
\label{fig:FBias}
\vspace{-0.2truecm}
\end{figure*}

A post-it note analogy is well illustrated on the example of graphene/h-BN/Co, where h-BN, an insulator of band gap $E_g \sim 6$~eV, has been widely
employed to improve transport, optical, and spin-dependent properties of graphene and other vdW materials~\cite{Drogeler2016:NL,Mak2016:NP}. 
The metallic Co dopes graphene 
and shifts the Dirac cone far below the Fermi level,  
but a large $E_\mathrm{ext}$ can return it back to the Fermi level. Remarkably, 
by comparing the results of $E_\mathrm{ext}= -0.32$ V/$\mathrm{\AA}$ and $-0.4$  V/$\mathrm{\AA}$ for the Dirac cone slightly below ($n$-doped) and slightly above ($p$-doped) 
$E_F$, respectively, the proximity-induced $P$ in graphene not only changes its magnitude, but also reverses its sign~\cite{Lazic2016:PRB}. Instead of the usual reversal 
of spin or magnetization by an applied magnetic field, this prediction suggests that a gate-control of spin reversal is possible. This proximity-induced reversal of spin polarization was
further corroborated in another first-principles study of heterostructures of graphene and Co separated by 1, 2, and 3 h-BN layers which was complemented by the
phenomenological electronic structure model~\cite{Zollner2016:PRB}. The proximity induced spin splitting of graphene reached $\sim 10$ meV for a single h-BN,  
decreasing 
in magnitude but altering sign as additional layers were inserted, similar to the spatial dependence of IEC. As expected, with more h-BN layers 
there is a decrease in the doping of graphene and the shift of the Dirac cone from the Fermi level~\cite{Zollner2016:PRB}.  

An early motivation to fabricate F/graphene junctions was stimulated by the prediction that graphene can provide effective spin filtering~\cite{Karpan2007:PRL} 
or replace a tunnel barrier, having the advantage of low resistance and a small number of defects~\cite{vantErve2012:NN,Dery2012:NN}. 
Resulting structures would be suitable for a robust spin injection or a large 
TMR. In contrast to that focus on
the ideally lattice-matched single-crystalline F/graphene structures required for effective spin filtering, recent experiments~\cite{Asshoff2017:2DM} suggest  
a different picture based instead on  vdW heterostructures~\cite{Geim2013:N}, formed without lattice-matched crystals. The role of graphene was
then viewed not as spin filter, but a source of spin-polarized carriers itself, arising from an interplay of doping by the F metal and the 
proximity-induced spin splitting in graphene, similar to what can be expected from Fig.~6(d).

Related transport experiments on vertical Co/ graphene/h-BN/NiFe junctions in Figs.~\ref{fig:FBias}(a)-(d) demonstrate that the bias-dependence 
of the measured MR can change both its magnitude and sign. From Julli{\'e}re's formula~\cite{Zutic2004:RMP}, 
$MR=2 P_\mathrm{Co}P_\mathrm{Py}/(1-P_\mathrm{Co}P_\mathrm{Py})$, where $P_\mathrm{Co/Py}$ is the DOS spin polarization of
the Co- or NiFe-proximitized graphene. In a simple model, $|P|=E_\mathrm{ex}/2|E_F|$, where at 10 K the proximity-induced exchange splitting at zero
bias is estimated in graphene to be $E_{ex} \sim 6$ meV~\cite{Asshoff2017:2DM}, 
 of the same order of magnitude, but larger than in heterostructures  of graphene or TMDs with 
insulating F from Figs.~8(f) or 9(c). MR is also present at 300 K, although reduced by $\sim 40$ \%, as expected from the thermal reduction of 
the effective spin polarization due to magnons~\cite{Shang1998:PRB,Garzon2005:PRL}.

The observed behavior supports the role of proximity effects leading to the spin-dependent DOS in graphene, put forth in the interpretation of Figs.~\ref{fig:FGate}(d)-(f).
However, rather than the gate-controlled  $E_\mathrm{ext}$, an applied bias creates a relative shift of the DOS with respect to the Fermi level. 
A change from $n$- to $p$-doped graphene is consistent with the sign reversal in the measured MR~\cite{Asshoff2017:2DM}. 
The absence of lattice matching between the metallic F and the adjacent graphene layers preserves the Dirac cone by suppressing the hybridization 
that would be expected for epitaxial graphene/F metal heterostructures [Fig.~11(b)]. The measured large bonding distances [recall Fig.~13(a)], 
it was 7a
$d_\mathrm{Co/Graphene}=3.9 \pm 0.6$ $\mathrm{\AA}$ 
and $d_\mathrm{Py/Graphene}=3.4 \pm 0.9$ $\mathrm{\AA}$, are consistent with the vdW bonding and thus effective gate/bias-controlled changes in the electronic structure~\cite{Lazic2016:PRB}.

In a lateral geometry from Fig.~\ref{fig:FBias}(e) which also employed Co/h-BN/graphene junctions, nonlocal measurements of spin injection and detection [recall Fig.~7] have shown a large bias-induced
differential spin injection and detection polarizations. These results reveal a striking behavior that the spin polarization can be reversed close to zero applied bias, Fig.~\ref{fig:FBias}.
A strong bias-dependence of spin-polarization qualitatively differs from the simple description of spin injection based on the equivalent resistor scheme and a linear $I-V$ characteristics.
However, a sign change of the spin polarization with bias was predicted in magnetic {\it p-n} junctions, distinguishing the cases of spin injection and extraction (reverse vs forward bias)~\cite{Zutic2002:PRL}.

Even though the authors' interpretation~\cite{Gurram2017:NC} of the observed results is not attributed to proximity effects, the effective fields obtained from the applied bias, 
while smaller than in Fig.~\ref{fig:FGate}(e), are not incompatible with magnetic proximity effects. By employing bilayer h-BN the coupling between Co and graphene 
is weaker than for single layer h-BN and thus there will be reduced doping effects and reduced required values of applied bias/field to bring Dirac cone back close to the Fermi level.

The most direct support for tunable magnetic proximity effects has been recently demonstrated in specially designed 1D edge contacts between Co and h-BN encapsulated 
graphene and measured gate-dependent nonlocal spin transport~\cite{Xu2017:P} similar to the geometry from Fig.~\ref{fig:FGate}(e). The 1D contacts, which have been previously
realized with nonmagnetic metals~\cite{Wang2013:S}, show a weaker coupling  between Co and graphene than in conventional 2D counterparts, which have enabled a lower applied $E_\mathrm{ext}$ than in Fig.~\ref{fig:FGate}(e), for the gate-controlled sign reversal of proximity-induced  spin polarization in graphene.

\subsection{3.4 Proximity with Antiferromagnets} 

Antiferromagnets (AFs) have recently attracted intense interest for a variety of spintronic and magnetoelectronic applications~\cite{MacDonald2011:PTRSA,Jungwirth2016:NN,Gomonay2017:PSS,Binek2005:JPCM,
Wang2014:PRL,Baltz2018:RMP}. 
Some of the attractive features include the absence of and insensitivity to stray magnetic fields and ultrafast dynamics arising from 
the fact that the precession frequencies are enhanced by  the exchange interaction~\cite{Kittel1951:PR}. The AF domain state can be manipulated 
by electric current in metallic AFs without macroscopic time-reversal symmetry~\cite{Zelezny2014:PRL}, while AF films with strong Dzyaloshinskii-Moriya 
interaction~\cite{Tsymbal:2011} can support topologically protected skyrmions which, in contrast to F skyrmions, move strictly along the electric 
current~\cite{Barker2016:PRL,Zhang2016:SR}. These features could potentially be exploited in memory and logic devices.
The N\'eel temperature of a thin AF layer may be enhanced beyond its bulk value when in proximity to another magnetic layer. For example, 
this has been observed for AF CoO next to ferrimagnetic Fe$_3$O$_4$~\cite{Zaag2000:PRL} or AF NiO~\cite{Borchers1993:PRL,Abarra1996:PRL}. 
A number of other examples have been reviewed in Ref.~\cite{Manna2014:PR}.

As mentioned above, proximity effects become essentially a bulk phenomenon in atomically 2D materials. To our knowledge, one material with in-plane AF ordering, 
FePS$_3$, has been obtained in ML form through mechanical exfoliation~\cite{Wang2016:2DM}. It is predicted that similar Mn-based compounds may have an AF 
ordering commensurate with the crystallographic (honeycomb) unit cell, which breaks macroscopic time-reversal symmetry and couples to the valley degree of freedom~\cite{Li2013:PNAS}.

\begin{figure}[ht]
\vspace{0.15cm}
\resizebox{8.7cm}{!}{\includegraphics{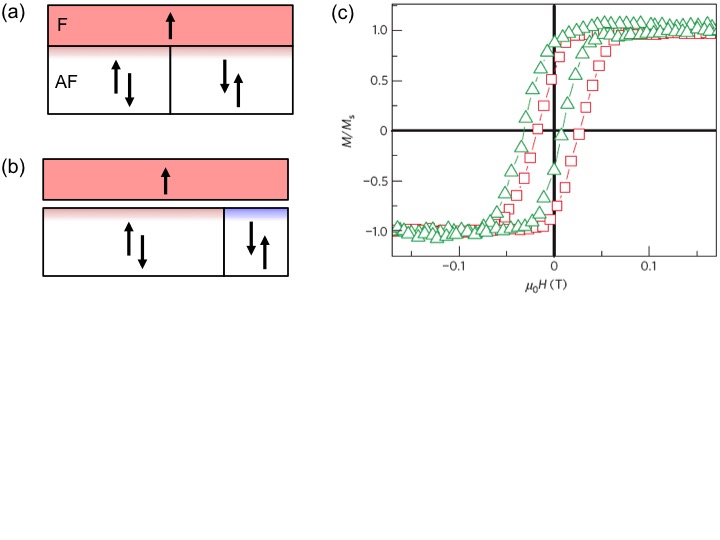}}
\vspace{-3.5cm}
\caption{\footnotesize{%
Exchange bias phenomenon. Panels (a) and (b) schematically illustrate the classification of exchange bias mechanisms into two types based on the (a) presence or (b) absence of 
macroscopic time-reversal symmetry in the bulk of AF. Arrows in AF: the sublattice {\bf M}. In 
(a), there is no imbalance in the AF domain occupations, and the exchange bias is due to the "frozen-in" proximity {\bf M} 
near the surface (red gradient coloring near the interface with F). In 
(b), field-cooling creates a preponderance of one AF domain; the two domains have opposite 
surface {\bf M} (red or blue gradient coloring), which is an equilibrium property that does not require F proximity. 
(c) Hysteresis loops of a Co/Pd multilayer interfaced with Cr$_2$O$_3$ (0001), with an opposite shift created using magnetoelectric annealing~\cite{He2010:NM}. 
Adapted with permission (c) from Ref.~\cite{He2010:NM}.}}
\label{fig:exbias}
\end{figure}

By analogy with ferromagnetic proximity, the incorporation of 
AF layers in vdW heterostructures may bring a wealth of opportunities for novel phenomena and applications. vdW interlayers like graphene can also mediate interlayer 
exchange coupling, which can be used to engineer synthetic antiferro- or ferrimagnetic heterostructures~\cite{Gargiani2017:NC}. 
It was shown that proximity 
effects in heterostructures combining layers of a magnetically doped topological insulator and AF CrSb can induce a modulation of the interfacial spin texture and,  at the same time, enhance the Curie temperature of the superlattice~\cite{He2017:NM}.

It is interesting to consider the role of the magnetic proximity effect in the exchange bias phenomenon, widely used for magnetically storing and sensing information~\cite{Nogues2005:PR}, 
which manifests itself in the shift of the hysteresis loop of F 
interfaced with AF, 
along the magnetic field axis. This shift requires that the macroscopic 
time-reversal symmetry is broken by the AF. Conceptually,
one can identify two qualitatively different mechanisms of this symmetry breaking, depicted in Fig.~\ref{fig:exbias}. 
In the first mechanism, which is possible with any AF material, 
the time-reversal symmetry is broken during magnetic field-cooling: The proximity exchange field from the F induces ${\bf M}$ in the AF near the surface,
which is subsequently ``frozen in.'' This {\bf M} is nonequilibrium, and, therefore, this conventional mechanism of exchange bias is often susceptible to the so-called 
training effect, as the successive hysteresis loop cycles tend to unfreeze the nonequilibrium ${\bf M}$ and reduce the exchange-bias field. The microscopic details of 
this mechanism are complicated and system-dependent~\cite{Nogues2005:PR}.

The second mechanism, which has only recently been understood~\cite{He2010:NM,Belashchenko2010:PRL}, requires an AF with broken 
macroscopic time-reversal symmetry. Such AF exhibit the magnetoelectric effect \cite{Fiebig2005:JPD} (Cr$_2$O$_3$ is a common example 
that has been extensively studied due to its relatively high N\'eel temperature) and, by virtue of their magnetic symmetry, have the following properties: 
(1) different AF domains are macroscopically distinguishable, (2) \emph{magnetoelectric} field-cooling \cite{Martin1966:IEEETM} can be used to 
favor one domain type over the other(s), thereby breaking the time-reversal symmetry throughout the whole bulk of the AF, (3) the surface of 
this material carries an \emph{equilibrium} {\bf M}, which is not destroyed by roughness~\cite{Andreev1996:JETP,Belashchenko2010:PRL}. Thus, a 
magnetoelectrically field-cooled AF of this kind creates an equilibrium exchange bias in a proximate F, as long as the magnetocrystalline 
anisotropy is high enough to prevent the switching of the AF domain state throughout its bulk. Being an equilibrium effect, such exchange bias 
usually does not exhibit the training effect.

The generic first mechanism of the exchange bias requires a nonequilibrium retention of the F proximity effect near the surface of the 
AF, while the second mechanism, typical for magnetoelectrics, does not require proximity effect at all.

Measurement of the anomalous Hall effect in a thin Pt overlayer has been turned into a detection technique for the surface magnetization~\cite{Andreev1996:JETP,Belashchenko2010:PRL} 
of AF chromia (Cr$_2$O$_3$)~\cite{Kosub2015:PRL}, which could become an essential ingredient in magnetoelectric memory devices~\cite{Kosub2017:NC,He2010:NM}.

\section{4. Spin-Orbit Proximity Effects} \label{Sec:SOPE}

\subsection{4.1 Interfacial Spin-Orbit Coupling} 

\begin{figure*}
\resizebox{18cm}{!}{\includegraphics{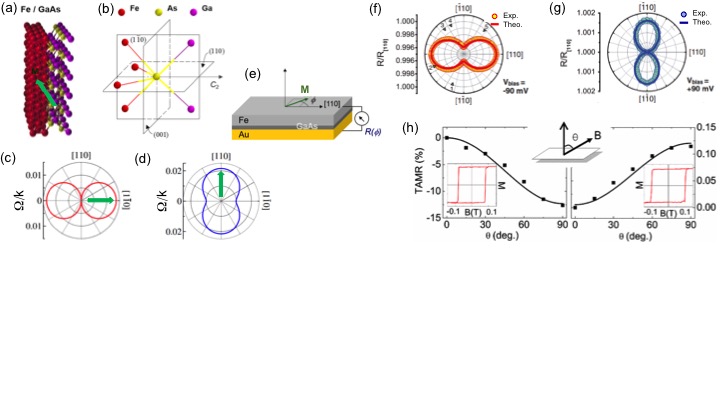}} 
\vspace{-4.6cm}
\caption{(a) Structure of a Fe/GaAs slab. (b) The nearest As neighbors at the Fe/GaAs interface. The interface has the symmetry of the point group $C_{2v}$, containing a $C_2$ rotation axis and two mirror planes (110) and $(1\bar{1}0)$. (c) Angular dependence in $\mathbf{k}$-space of the amplitude, $\Omega=|\mathbf{\Omega}(\mathbf{k})|$, of the ISOC field for {\bf M} along the GaAs $[1\bar{1}0]$ direction (green arrow). 
(d) Same as in (c) but for $\mathbf{M}$ along the [110] direction~\cite{Gmitra2013:PRL}. (e) Experimental setup for in-plane TAMR. $\mathbf{M}$, is rotated in the plane of the magnet, the tunneling resistance $R$ is measured as a function of $\phi$ and normalized to its $\phi = 0$ value, $R_{[110]}$. Experimental results for bias voltages of -90~meV and 90~meV are shown in (f) and (g), respectively~\cite{Moser2007:PRL}. 
(h) Angular dependence of the TAMR in the out-of-plane configuration. Left and right panels correspond to CoPt/AlO$_x$/Pt and Co/AlO$_x$/Pt tunnel junctions, respectively. The presence of an extra Pt layer with strong SOC yields a TAMR in CoPt/AlO$_x$/Pt  two orders of magnitude larger than in Co/AlO$_x$/Pt. The insets show magnetization measurements in out-of-plane magnetic fields~\cite{Park2008:PRL}. 
Adapted with permission (a)-(d) from Ref.~\cite{Gmitra2013:PRL}, (e)-(g) from Ref.~\cite{Moser2007:PRL}, (h) from Ref.~\cite{Park2008:PRL}.}
\label{fig:ISOC}
\end{figure*}

Proximity effects are commonly realized by bringing together two or more materials leading to the formation of interfaces between them. The inherent lack of inversion symmetry 
at interfaces yields the formation of interfacial spin-orbit coupling (ISOC). Therefore, whether negligible or not, ISOC is intrinsically related to proximity effects. 
ISOC can also appear at a surface, which can be understood as the interface between a given material and vacuum.
As in our discussion of SOC in Sec.~3.2, 
for bands with a 2D representation, the corresponding Hamiltonian is given by SOC field $\mathbf{\Omega(k)}$, 
$H_\mathrm{SO}=\mathbf{\Omega}(\mathbf{k}) \cdot \bm{\sigma}$. A simple case is a so-called Rashba SOC with $\mathbf{\Omega(k)}=(\alpha k_y, -\alpha k_x)$, responsible for
chiral spin textures~\cite{Zutic2004:RMP,Fabian2007:APS,Winkler:2003,Rashba1960:SPSS}.

The ISOC contains information about the interface-induced symmetry reduction of the individual bulk constituents. An instructive example is an Fe/GaAs junction, where the cubic and $T_d$ symmetries of bulk Fe and GaAs, respectively, are reduced to $C_{2v}$ in the heterostructure, as shown in Figs.~15 
(a) and (b). The formation of ISOC fields is therefore crucial for qualitatively new phenomena, absent or fragile in the bulk, such as the tunneling anisotropic 
MR~\cite{Gould2004:PRL,Moser2007:PRL}, crystalline anisotropic 
MR~\cite{Hupfauer2015:NC}, magneto-anisotropic Andreev reflection~\cite{Hogl2015:PRL}, 
SO torques~\cite{Garello2013:NN,Haney2013:PRB,Brataas2014:NN,Kim2017:PRB}, 
skyrmions~\cite{Fert2013:NN,Jiang2015:S,Nayak2017:N,Gungordu2016:PRB}, tunneling anomalous and planar Hall effects~\cite{Tarasenko2004:PRL,Vedyayev2013:APL,%
Matos-Abiague2015:PRL,Dang2015:PRB,Scharf2016:PRL,Chiba2017:PRB}. 

Since the ISOC is present only in the vicinity of the interface, its effects can be controlled electrically by gate voltage or an applied external bias capable of pushing the carriers wavefunction 
into or away from the interface. ISOC can also be controlled magnetically.
Calculations for an Fe/GaAs slab have revealed that when the Fe-GaAs 
hybridization is strong enough, the emergent ISOC strongly depends on the $\mathbf{M}$ orientation in the Fe layer \cite{Gmitra2013:PRL}, as illustrated in 
Figs.~15(c) and (d). 

\subsection{4.2 Tunneling Anisotropic Magnetoresistance} 

Tunelling anisotropic magnetoresistance (TAMR) is the dependence of the tunneling current in a tunnel junction with only \emph{one} magnetic electrode on the spatial orientation of its magnetization \cite{Gould2004:PRL}. 
For an in-plane rotation of $\mathbf{M}$ depicted in Fig.~15, 
we can define it as the normalized resistance difference, $\mathrm{TAMR}=(R(\phi)-R_{[110]})/R_{[110]}$. 
TAMR appears because the electronic structure depends on this orientation, due to SOC. In the context of proximity effects, the electronic structure near the magnetic interface is of interest. In particular, a surface or an interface can host pure or resonant bands. The Fe (001) surface provides a well-known example \cite{Stroscio1995:PRL}. In the presence of SOC, the dispersion of these states depends on the $\mathbf{M}$ orientation  \cite{Chantis2007:PRL}. As a result, the tunneling conductance, which, in a crystalline junction, is very sensitive to the transverse wave vector, develops both out-of-plane and in-plane MR, 
whose angular dependence reflects the crystallographic symmetry of the interface. For example, the TAMR inherits the $C_{4v}$ symmetry for the Fe (001) surface \cite{Chantis2007:PRL} and the reduced $C_{2v}$ symmetry for the Fe(001)/GaAs interface \cite{Moser2007:PRL}. In the latter case, the SOC originating in GaAs affects the electronic structure at the magnetic interface, which can be viewed as a 
SO proximity effect, appearing also with topological insulators~\cite{Marmolejo-Tejada2017:NL}.
A similar effect, combining low crystallographic symmetry of the interface with SOC, manifests itself in the angular dependence of the SO torque in F/heavy-metal 
bilayers~\cite{MacNeill2017:NP}.

\subsection{4.3 Crystalline Anisotropic Magnetoresistance} 

\begin{figure*}
\resizebox{18cm}{!}{\includegraphics{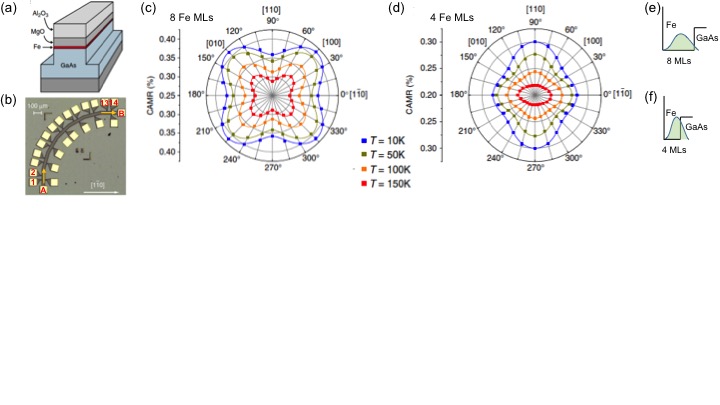}}
\vspace{-5.7cm}
\caption{(a) Sketch of the structure used to measure the CAMR. (b) Optical micrograph of the quadrant type sample. 
The mesa, defines the transport channel and current flow from contact A to B. Lateral contacts are used to measure simultaneously longitudinal voltage drops along seven different crystallographic directions. For example, the contacts 1 and 2 (13 and 14) measures the voltage drop for a current along the [110] ($[1\bar{1}0]$). The CAMR coefficient IS the contrast between the  longitudinal voltages $U_{\rm max}$ and $U_{\rm min}$, CAMR$=[U_{\rm max}(\theta)-U_{\rm min}(\theta)]/[U_{\rm max}(\theta)+U_{\rm min}(\theta)]$,
measured when $\mathbf{M}$ is parallel and perpendicular to the current direction. (c) CAMR as a function of the direction of the current flow ($\theta$, is measured with respect to the GaAs $[1\bar{1}0]$ direction) at different temperatures for a sample with 8 monolayers (MLs) of Fe. (d) Same as in (c) but for a sample with 4 MLs of Fe. The symbols are experimental data while solid lines are fits using a phenomenological model.
(e) and (f) As the number of Fe MLs decreases the hybridization between Fe-like and GaAs-like states increases due to an enhancement of the wavefunction penetration into the undoped GaAs region. This leads to the enhancement of the ISOC effect on the CAMR, reducing its symmetry from four-fold to two-fold.
Adapted with permission (a)-(d) from Ref.~\cite{Hupfauer2015:NC}.
}
\label{fig:CAMR}
\end{figure*}

The anisotropic MR (AMR) accounts for the difference in the resistances measured when the magnetization is parallel and perpendicular to the current flow~\cite{Zutic2004:RMP}. 
SOC couples the carrier momentum defined with respect to given crystallographic axes to its spin and can lead to the so-called crystalline AMR (CAMR) effect which refers to the anisotropy 
of the AMR with respect to the direction of the current~\cite{Rushforth2007:PRL,Ranieri2008:NJP,Hupfauer2015:NC}. 

When the SO proximity effect is negligible, the nature of the CAMR is determined by the bulk SOC as, for example, in (GaMn)As layers \cite{Rushforth2007:PRL,Ranieri2008:NJP}. 
However, in ultra-thin films the SO proximity effect due to ISOC can even dominate over the bulk SOC contribution (see Fig.~\ref{fig:CAMR}). This has been experimentally demonstrated 
by measuring the CAMR in ultra-thin films of epitaxial Fe/GaAs(001) \cite{Hupfauer2015:NC}. Figures \ref{fig:CAMR}(c) and (d) show polar plots of the CAMR as a function of the current direction with respect to the GaAs $[1\bar{1}0]$ crystallographic axis, for the cases of 8 and 4 monolayers (MLs) thick Fe, respectively. The presence of both bulk-like and interfacial SOC yields the overall two-fold symmetry observed in the measured CAMR. However, as the thickness of Fe layers decreases from 8 to 4 MLs, the CAMR symmetry dominated by four-fold-Fe bulk like SOC evolves into a two-fold $C_{2v}$ symmetry dominated by the ISOC. Therefore, the reduction of the CAMR symmetry represents a direct evidence of the SO proximity effect on the transport properties of ultra-thin Fe, due to the presence of the nearby undoped GaAs. An additional signature of the SO proximity effect is the reorientation of the CAMR main symmetry axes from $([100],[010])$ to $([1\bar{1}0],[110])$ when decreasing the Fe thickness [see Figs.~\ref{fig:CAMR}(c) and (d)].
The CAMR measurements in Fe/GaAs showed that the strength of the SO proximity effect can be increased by decreasing the Fe thickness down to 4 MLs, suggesting that the effect could be further increased if 2D crystals are considered.

\subsection{4.4 Graphene/Transition Metal Dichalcogenide Hetrostructures}

Composed of carbon atoms, a light element, graphene possesses a rather weak intrinsic SOC, allowing for a long spin-relaxation length and spin lifetime \cite{Zomer2012:PRB,Neumann2013:S,Guimaraes2012:NL,Guimaraes2014:PRL,Han2014:NN,Drogeler2016:NL}.  While this may be advantageous for 
efficient spin transport, the intrinsically small SOC poses challenges for controlling spins and modulating spin currents by electrical means thus complicate the realization of graphene-based spintronic 
spin switches and transistors relying on SOC. 
However, we recall that an alternative implementation could utilize tunable magnetic proximity effects in graphene~\cite{Lazic2016:PRB}, discussed also in Sec.~5.4.

In order to enlarge its SOC, graphene can be functionalized by adding other atoms. For example,  
light atoms like hydrogen could enhance the SOC-induced energy gap by an order of magnitude, from about 24 $\mu$eV in pristine graphene \cite{Gmitra2009:PRB,Abdelouahed2010:PRB} to about 0.2 meV in semi-hydrogenated graphene \cite{Gmitra2013:PRL}. The SOC can, in principle, 
be further enlarged by adding heavy atoms but it comes at the price of stronger changes in the local electronic structure
and an increase of undesired disorder effects.

\begin{figure*}
\resizebox{18cm}{!}{\includegraphics{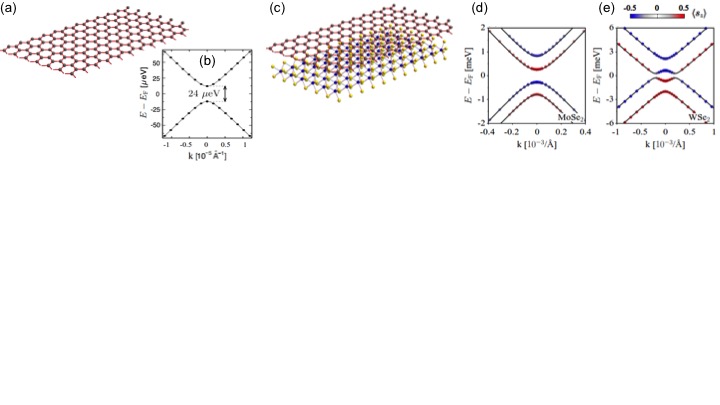}}	
\vspace{-6.8cm}
\caption{Pristine graphene (a) and its band structure (b). The weak intrinsic SOC leads to a rather small energy gap (about 24~$\mu$eV) opening in the Dirac cone, without breaking the spin degeneracy. The SO proximity effect in a graphene/TMD heterostructure (c) enhances the overall strength of SOC in the graphene layer, as evidenced in the graphene/MoSs$_2$ band structure (d). In addition to the intrinsic SOC, the graphene/TMD heterostructures exhibit Rashba and PIA SOCs, which results in the breaking of the spin degeneracy (the line colors indicate the spin projection along a direction perpendicular to the layers) and in some structures such as graphene/WSe$_2$ can even lead to a band inversion (e). Adapted with permission (b) from Ref.~\cite{Gmitra2009:PRB}, (d) and (e) from Ref.~\cite{Gmitra2016:PRB}.}
\label{fig:G-TMD}
\vspace{-0.2cm}
\end{figure*}

Another way of enhancing the SOC in graphene is to use the SO proximity effect in a heterostructure coupling graphene to a material containing heavy elements. 
Such an approach has been investigated both theoretically and experimentally in graphene/TMD 
systems~\cite{Jin2013:PRB,Zhang2014:PRL,Lee2015:ACSN,Lin2017:NL,Benitez2017:NP,Ghiasi2017:NL}, indicating an increase in the SOC-induced gap by two to three orders of magnitude compared to pristine graphene.

Unlike graphene,
TMDs exhibit a strong SOC (due to the $d$ orbitals of the heavy metal atoms). In the ML TMDs the lack of a center of inversion symmetry leads to the 
coupling between the spin- and valley ($K$/$K'$ points)-degrees of freedom and pins the spins of electrons with opposite momenta to opposite 
perpendicular-to-the-plane directions. On the other hand however, the carrier mobilities in graphene are much higher than in TMDs. Therefore, the 
SO proximity effect in graphene/TMD heterostructures represents a promising approach for the development of spintronic devices integrating the 
exceptional transport properties of graphene with the SOC-mediated electric control of spins. Indeed, TMDs through proximity effect enhance the 
SOC in graphene (see Fig.~\ref{fig:G-TMD}) by nearly three orders of magnitude, allowing for the realization of the spin Hall effect~\cite{Avsar2014:NC}, 
and weak antilocalization~\cite{Wang2015:NC}. In Sec.~5.3 we discuss how the graphene/TMD structures are used to implement 
spin switches and electric gate control of spin current~\cite{Yan2016:NC,Dankert2017:NC}.
In addition to the Rashba-like SOC resulting from the lack of structure inversion symmetry, the pseudospin inversion asymmetry (PIA) in 
graphene/TMD structures gives rise to an extra contribution to the SOC. Theoretical estimates of the strength of the PIA-induced SOC in various 
graphene/TMD heterostructures have been reported in Ref.~\cite{Gmitra2016:PRB}.

Our understanding that {\em equilibrium} proximity effects in atomically-thin materials  have also important {\em nonequilibrium} implications 
[recall the discussion of Fig.~6(d)] is 
verified in the case of SO proximity. With highly-anisotropic SOC in TMDs [Sec.~3.2, Fig.~9(d)], 
we expect that the proximity-induced SOC will also be anisotropic in graphene. 
Through 
nonlocal spin transport measurements 
for F1/graphene/WS$_2$/F2 junction, the observed spin lifetime in graphene 
was highly anistotropic with the direction of an applied magnetic field which determines the spin precession~\cite{Benitez2017:NP}. 
Consistent with the native SOC anisotropy in TMDs, even at 300 K there was a ten-fold increase in the spin lifetimes for the out-of-plane 
spins as compared to the in-plane spins~\cite{Benitez2017:NP}.
These results suggest that, through SO proximity, the spin-valley coupling of 
TMDs was imprinted in graphene.

Strong SOC and spin-valley coupling in TMDs implies that the emitted or absorbed light have valley-selective helicity~\cite{Xiao2012:PRL}
motivating the proposal to use valley polarization generated by circularly polarized light in TMD 
to optically inject spin in the nearby graphene where with only a weak SOC a direct optical spin injection
would be ineffective~\cite{Gmitra2015:PRB}. This scheme, demonstrated experimentally~\cite{Luo2017:NL,Avsar2017:ACSN} provides 
another example of how vdW heterostructures with regions of different SOC strengths could enable useful functionalities. 

A similar scenario was earlier proposed for spin injection and detection in Si~\cite{Zutic2006:PRL}, sharing with graphene desirably 
long spin relaxation times and spin diffusion lengths as well as a weak SOC which precludes effective optical spin injection and detection. 
However, a Si-based heterostructure with a direct band gap semiconductor of larger SOC, such as GaAs, 
could overcome this difficulty. Through spin diffusion, a circularly-polarized light illuminating GaAs could enable 
optical injection of spin into the nearby Si or, alternatively, spin injected in Si could be optically detected in the nearby GaAs
through the circular polarization as luminescence, as confirmed experimentally~\cite{Jonker2007:NP,Jonker:2011}.

The SO proximity effect in graphene/TMD heterostructures may also lead to the emergence of topological phases. Theoretical calculations indicate that band inversion can occur in graphene/WS$_2$ and graphene/WSe$_2$ [see Fig.~\ref{fig:G-TMD}(d)], causing the formation of topologically protected helical edge states and the realization of the quantum spin Hall effect \cite{Wang2015:NC,Gmitra2016:PRB}. 
The transition from the inverted-band quantum spin Hall phase to a direct-band phase exhibiting the valley Hall effect could be controlled by modulating the strength of the SO proximity effect with a gate voltage \cite{Alsharari2016:PRB}.

\section{5. Applications}\label{Sec:APP}

\subsection{5.1 Overview} 

With our focus on tailoring spin-dependent properties
of materials using proximity effects, the resulting 
applications can be  mostly viewed in the context of spintronics, but  
not necessarily limited to magnetic hard drives or MRAM in which the
use of proximity effects through exchange bias (recall Sec.~3.4) is
already commercialized. 
Chosen examples serve two purposes: (i) to examine opportunities in which proximity effects
could complement or replace other schemes for realizing spintronic devices, (ii) to stimulate exploring 
different systems where proximity effects could enable novel applications.
For example, magnetic proximity effects could allow us to rethink not only how to process information 
and implement low-power spin logic, but also how to seamlessly integrate nonvolatile memory and logic. 
On the other hand, using spin for transferring information can be boosted by proximity effects including
a novel class of spin lasers.

We also note a broader scope of possible applications.  
Superconducting proximity effects were the first to enable commercial applications, building on the discovery of the Josephson effect~\cite{Josephson1962:PRL}.
It relies on proximity-induced superconductivity across a normal region sandwiched between 
two superconductors. Once the voltage is applied across this device a dissipationless supercurrent 
flows. Such a Josephson junction is the key element of  a superconducting quantum interference device (SQUID)~\cite{Tinkham:1996} which provides extremely 
sensitive detection of magnetic fields (as small  10$^{-17}$ T) finding its use from the studies of biological systems and magnetic resonance imaging, 
to the detection of gravitational waves~\cite{Romani1982:RSI,Goryachev2014:PRD}.

The interest in superconducting proximity effects has been recently extended to fault-tolerant quantum computing with  
exotic quasiparticles 
known as the Majorana Fermions or Majorana bound states which are their own antiparticles~\cite{Elliot2015:RMP}. 
A pair of these spatially separated Majorana states enables a peculiar realization of an electron, making them robust against local perturbations 
that are detrimental to other quantum computing implementations. 
Unlike the exchange of two electrons which lead to an overall sign change of their 
wavefunction, the exchange of two Majorana bound states effectively acts as a matrix, transforming their wavefunction into a new state, therefore 
implementing a quantum gate~\cite{Aasen2016:PRX,DasSarma2015:QI,Kitaev2003:AP}.

\subsection{5.2 Spin Interconnects} 

\begin{figure}[ht]
\vspace{0.1cm}
\resizebox{8.1cm}{!}{\includegraphics{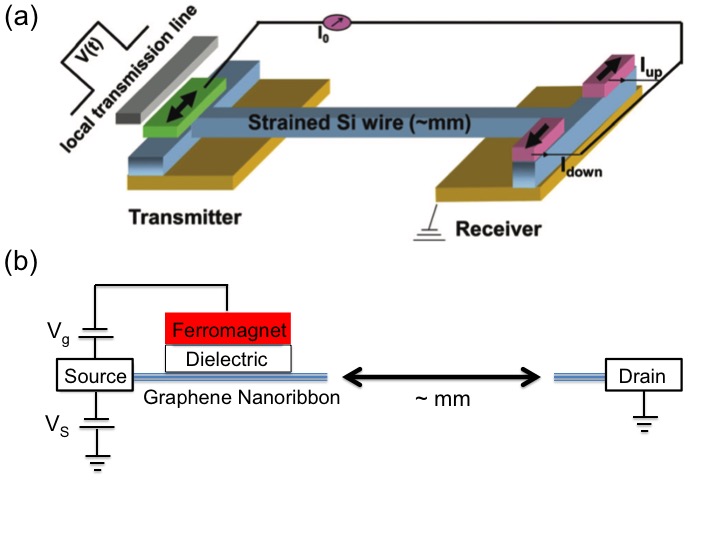}}
\vspace{-1cm}
\caption{\footnotesize{%
A spin-based communication scheme. The information is encoded by modulating the spin polarization at a constant charge current.
(a) The transmitter relies on the time-dependent magnetic field to reverse {\bf M}
in the spin injector. The receiver splits the current into two paths (right contacts) and detected logical ``1" or ``0" based on
the predominance of spin-up (I$_\mathrm{up}$ $>$I$_\mathrm{down}$) or spin-down (I$_\mathrm{up}$ $<$ I$_\mathrm{down}$) currents. 
Not drawn to scale, the Si channel is far longer than any of the dimensions of the transmitter and receiver circuits~\cite{Dery2011:APL}.
(b) An alternative realization. The transmitter employs a tunable magnetic proximity effect~\cite{Lazic2016:PRB} to modulate the spin polarization
in the graphene nanoribbon. Adapted with permission (a) from Ref.~\cite{Dery2011:APL}.
}}
\label{fig:coma}
\vspace{-0.4cm}
\end{figure}

Conventional charge-based metallic interconnects are becoming the key obstacle in the continued scaling of integrated circuits. 
In on-chip communications, signals are transmitted via metallic wires, modeled as transmission lines with the voltage and current 
being distance and time-dependent. In addition to their drawbacks such as dynamical crosstalk between wires, RC bottlenecks, and 
electromigration, these interconnects are also 
the main source of energy consumption~\cite{ITRS:2015,Miller2017:JLT,Hilbert2011:S}.
Resulting effects become increasingly acute with reducing the spacing between adjacent wires and with increasing the modulation frequency. 
To solve these problems, alternatives are considered~\cite{ITRS:2015,Miller2017:JLT}. 
One of them   
relies on spin-based on-chip data communication, shown in Fig.~\ref{fig:coma}. 

The idea is to modulate the electrons' spin polarization 
of a constant current in wires of group IV materials (Si, Ge, or graphene). The intrinsic limit to the channel length is set by the decay of spin information. 
When electrons drift at nearly the saturation velocity (e.g., 10$^7$ cm/s in silicon), this length scale readily reaches 1 mm at room temperature~\cite{Dery2011:APL}. 
It is potentially longer in strained Ge~\cite{Li2012:PRB} or in high-mobility graphene nanoribbons. These length scales are already more than sufficient for on-chip 
interconnects in modern integrated circuits. Importantly, the constant-level current means elimination of crosstalk problems: the spin signal does 
not interfere with spin signals in wires similar to the one shown in Fig.~\ref{fig:coma}(a). 
Since this feature is independent of the wire density, the intrinsic limit 
to the information bandwidth density is expected to be orders of magnitude higher than what is currently feasible in metallic interconnects. 
For example, a very large bandwidth of 1000 Tbit/(cm$^2$s) can be supported with a Joule heating of $\sim$1 Watt caused by constantly driving the 
current in the interconnects~\cite{Dery2011:APL}.

An alternative realization of spin interconnects is possible using gate-tunable magnetic proximity effects~\cite{Lazic2016:PRB} to modulate spin 
polarization in graphene or graphene nanoribbons, depicted in Fig.~\ref{fig:coma}(b). The appeal of graphene and its nanoribbons comes from an ultrahigh mobility which can 
reach $\sim$ 10$^5$ cm$^2$/Vs at 300 K, while 
the nanoribbons can be fabricated to be narrower than 10 nm~\cite{Wang2008:PRL}. 
Applying the gate voltage, $V_g$, to modulate the proximity-induced spin polarization in a graphene nanoribbon, can alter the Fermi level and the constant  current condition.  It is therefore  important to include a compensating source voltage, $V_S$, to retain the constant current.  
The compensating source voltage modulation is local in the transmitter side and does not affect the constant charge current along the wire.
For the detection, not shown in Fig.~\ref{fig:coma}(b), different schemes 
are possible.  For example, as suggested for Si spin interconects~\cite{Chang2015:IEEETM}, 
one  can employ a spin transfer torque~\cite{Tsymbal:2011,Ralph2008:JMMM,Maekawa:2012}, induced by the spin current from a nanoribbon. 

\begin{figure*}[h]
\resizebox{18cm}{!}{\includegraphics{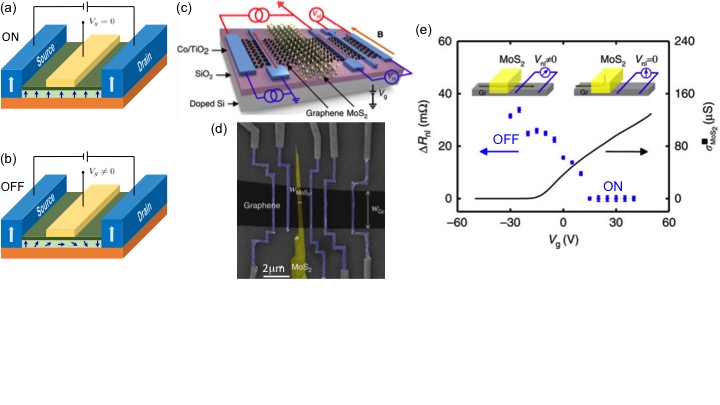}}
\vspace{-3.2truecm}
\caption{The Datta-Das spin FET is composed of F source and drain connected through a 2D electron gas as a transport channel. The top gate controls  the Rashba SOC in the transport channel. Electrons injected with momentum 
parallel to the transport undergo spin precession in a transverse SO field. 
(a) If $V_g =0$ the SOC strength vanishes and the spin does not precess, allowing electrons to enter into the drain (ON state). (b) At a certain $V_g$ the spin precesses by $\pi$ and the electron bounces back, increasing the channel's resistance (OFF state). 
(c) A sketch of a graphene/MoS$_2$ based spin field-effect switch.  A DC current is injected into graphene from a Co electrode across a TiO$_2$ barrier and a non-local voltage, $V_\mathrm{nl}$, is measured by a second Co electrode while sweeping the magnetic field B. The red- and blue-colored circuit diagrams: measurement configurations in the reference graphene lateral spin valve (LSV) and the graphene/MoS$_2$ LSV. 
(d) Scanning electron microscope image of the device. (e) Experimental demonstration of the ON/OFF state of the spin signal, $\Delta R_\mathrm{nl}$ (blue circles), by gate voltage, $V_g$. The black solid line is the MoS$_2$ sheet conductivity as a function of $V_g$. The insets show schematically the spin current path (green arrow) in the OFF state (left inset) and the ON state (right inset) of MoS$_2$. (c)-(e) reprinted with permission from \cite{Yan2016:NC}.}
\label{fig:SpinFET}
\end{figure*}

\subsection{5.3 Towards Spin Transistors} 

The so-called spin field effect transistor (FET) proposed by Datta and Das~\cite{Datta1990:APL} is essentially a three-terminal gate-modulated spin switch. 
As depicted in Figs.~19(a) and (b), F source and drain of the device have parallel {\bf M} and the current between them is modulated by the degree of 
spin precession, which is caused by the gate controlled Rashba SOC strength. Despite its conceptual simplicity, the realization of the Datta-Das spin FET has 
awaited 20 years~\cite{Koo2009:S} when it was demonstrated at $T=1.8$ K.

While a weak SOC in graphene makes it an excellent spin transport channel with long spin diffusion 
length~\cite{Drogeler2016:NL}, the same property poses a challenge for electrical 
SOC modulation of spin signal and implementing a spin switch. This difficulty was recently overcome in a lateral spin valve (LSV) based on graphene/few-layer 
MoS$_2$ heterostructure~\cite{Yan2016:NC} 
by exploiting a different mechanism for a spin switch. As shown in 
Figs.~19(c) and (d), 
the graphene/MoS$_2$ spin field-effect switch uses  F tunneling contact as a source for injecting spins into graphene and  F drain contact as a nonlocal spin detector.  A much stronger 
SOC and a moderate mobility of MoS$_2$ yields the spin diffusion length of only 20 nm, about two orders of magnitude smaller than in graphene~\cite{Dankert2017:NC}.  

In addition to the proximity-enhanced SOC in graphene, the dominant effect on spin transport is a gate-controlled MoS$_2$ sheet conductivity which changes 
by six orders of magnitude thereby changing the absorption of spins from the graphene channel as measured by the nonlocal MR, $\Delta R_\mathrm{nl}$~\cite{Yan2016:NC}, 
shown in Fig.~\ref{fig:SpinFET} %
19(e). For negative $V_g$, a small MoS$_2$ sheet conductivity forces the spin current to flow through the graphene 
channel and yields a larger nonlocal spin signal and thus a larger $\Delta R_\mathrm{nl}$ corresponding to the OFF state. With positive $V_g$ and a large sheet 
conductivity, the spin current is absorbed from graphene into MoS$_2$ which strongly reduces the spin signal due to a much smaller spin diffusion length in MoS$_2$ 
yielding a smaller $\Delta R_\mathrm{nl}$ in the ON state. The resulting difference in the current path and ON/OFF switch effectively selects between the small 
and large SOC.  While the spin switch mechanism was observed up to 200 K, this is not a significant limitation. 
The same principle was subsequently used 
to realize the spin switch in graphene/MoS$_2$ based LSV even at room temperature~\cite{Dankert2017:NC}.

Apart from its single layer version, bilayer graphene also possesses 
very good spin transport properties \cite{Yang2011:PRL,Avsar2016:AM}. It may have 
some technological advantages because it allows a more precise control of the chemical potential than in a single layer graphene. 
Therefore bilayer graphene/TMD hybrid structures are also promising  for the realization of spin FETs. 
A bilayer graphene/WSe$_2$ spin FET has recently been theoretically proposed \cite{Gmitra2017:PRL}. The device operates by gate tuning the spin relaxation time. 
The field-effect variation of the spin relaxation time in bilayer graphene on WSe$_2$ was estimated to be 4 orders of magnitude \cite{Gmitra2017:PRL}, providing  opportunities for a sizable 
modulation of the spin signal and a large contrast between the ON and OFF states.

\vspace{0.3cm}
\begin{figure*}
\resizebox{17.8cm}{!}{\includegraphics{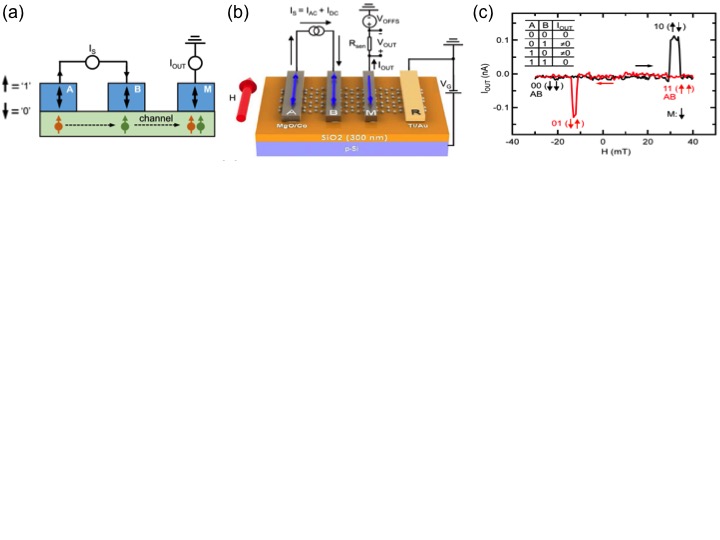}}
\vspace{-9.3cm}
\caption{
(a) Schematic of XOR magnetologic-gate device. 
$A $, $B$, and $M$ are F electrodes on top of a spin-transport channel. Input logic 1 and 0 are the two {\bf M} 
directions along the easy axis of the electrodes. $I_S$ injects spins through inputs, A and B. $I_\mathrm{out}$ is the logic output signal. (b) Device structure and measurement setup. 
$A$, $B$, and $M$ are MgO/Co electrodes. The spin channel is a single-layer graphene. $R$ is Ti/Au 
reference electrode used as ground point. $I_\mathrm{out}$ and $V_\mathrm{out}$ are the
current and voltage signal. $R_\mathrm{sen}$ is a variable resistor. $V_\mathrm{offs}$ 
is an ac voltage source. External magnetic field $H $ is applied to the easy axis of the electrodes. 
 (c) $I_\mathrm{out}$ measured as a function of $H$. Black (red) curve: H sweeps upwards (downwards). Vertical arrows: the 
 {\bf M} states of $A$ and $B$. Inset: truth table of XOR logic 
operation~\cite{Wen2016:PRA}. Adapted with permission from \cite{Wen2016:PRA}.
}
\label{fig:MLG}
\end{figure*}

Even though in this spin switch realization the proximity-induced SO in graphene was not the dominant effect (its presence was consistent with the reduced spin signal 
as compared to the graphene LSV without MoS$_2$), various implementations of  spin switches dominated by proximity effects are feasible, as  can be inferred 
from the gate-controlled magnetic proximity effects~\cite{Lazic2016:PRB}. 

Apart from its single layer version, bilayer graphene also 
possesses very good spin transport properties \cite{Yang2011:PRL,Avsar2016:AM}. It may have 
some technological advantages because it allows a more precise control of the chemical potential than in a single layer graphene. 
Therefore bilayer graphene/TMD hybrid structures are also promising  for the realization of spin FETs. 
A bilayer graphene/WSe$_2$ spin FET has recently been theoretically proposed \cite{Gmitra2017:PRL}. The device operates by gate tuning the spin relaxation time. 
The field-effect variation of the spin relaxation time in bilayer graphene on WSe$_2$ was estimated to be 4 orders of magnitude \cite{Gmitra2017:PRL}, providing  opportunities for a sizable 
modulation of the spin signal and a large contrast between the ON and OFF states.

\subsection{5.4 Spin Logic} 

Spintronic applications commonly employ magnetoresistive effects in which the resistance of a device can be changed by changing its {\bf M}. The nonvolatility of 
F is particularly suitable for magnetically storing or sensing information as given {\bf M} is preserved even in the absence of a power supply. However, beyond the success of magnetic hard disks and MRAM,  an outstanding challenge remains to employ such nonvolatility of ferromagnets as a means to seamlessly 
integrate memory and spin logic~\cite{Zutic2004:RMP,Sugahara2010:PIEEE}.

This tantalizing prospect offers also a paradigm change to overcome the inherent limitations of the widely employed logic circuits based on the von Neumann architecture. The design of such logic circuit relies on the central processing units connected by a communication channel to memory. While the bottleneck induced by data transfer across that channel can be alleviated by reducing the feature size of devices, it cannot be removed. Such bottlenecks are particularly obvious for data-intensive applications, where most of the actions involve accessing or checking data (rather than doing complex computation). Network routers are a classical example where the Internet Protocol address is compared with a list of patterns to find a match. Conventional CMOS implementation of such circuits suffers from scalability issues, making them ineffective for larger search problems that are important to contemporary tasks~\cite{Dery2012:IEEETED}. 

An initial proposal for a seamless integration of memory and logic using spin accumulation in Fe/GaAs lateral spin valves to implement magnetologic gates (MLGs)~\cite{Dery2007:N} has been subsequently extended to 
F/graphene junctions~\cite{Behin-Aein2010:NN,Dery2012:IEEETED}. A detailed circuit simulation for a MLG-based search engine which employs graphene for the spin propagation channel and CoFe and Py as hard and soft F regions, respectively, suggests its superior performance compared with optimized 32-nm CMOS counterpart designs~\cite{Agrawal2008:IEEEVLSI}. Other device advantages are 
associated with a related proposal of all-spin logic~\cite{Behin-Aein2010:NN}. 

The feasibility of such schemes for spin logic was boosted by the room temperature demonstration of the MLG built on 
graphene~\cite{Wen2016:PRA}. This MLG, depicted in Fig. 20, consists of three F electrodes contacting a single-layer graphene spin channel and relies on spin injection and spin transport in the graphene layer. 
The {\bf M} directions of the first two F electrodes ($A$, $B$) represent the logic inputs (0 and 1), and spin injection from these input electrodes generates a current through the third F electrode ($M$) which represents the logic output. 

A limitation of the current MLG implementation is the presence of an applied magnetic field, required to perform the {\bf M} switching. However, as discussed on the example of spin interconnects, an alternative realization could be provided by gate-tunable magnetic proximity effects.  Unlike the case of spin interconnects, for MLGs a constant charge current it is not 
required.
Material optimization should focus on moderate doping effects in the graphene channel such that the Dirac cone remains close to the Fermi level. This is an important prerequisite for the reversal of proximity-induced spin polarization at the values of external electric fields still attainable with conventional gating, rather than the much slower ion-liquid gating which allows for very large fields of almost 1 V/\AA~\cite{Mannhart1993:APL,Hebard1987:IEEETM}.

Experimental support for such a reduced doping and the Dirac cone close to the Fermi level has been provided by carefully designed 1D Co edge
contacts to h-BN encapsulated graphene to 
enable a gate-controlled reversal of the proximity induced spin polarization in graphene~\cite{Xu2017:P}.  
Alternatively, in 2D contacts depicted in Fig.~12(a), doping effects of a metallic F region could be compensated by placing another material with a suitable work function on the side of graphene opposite to the F region. 

\subsection{5.5 Spin Lasers} 

Lasers are ubiquitous in daily life with their applications including high-density optical storage, printing, optical sensing, display systems, and 
medical use~\cite{Chuang:2009,Coldren:2012,Michalzik:2013}.  To overcome the challenges of the continued Moore's law scaling discussed in 
Sec. 5.2, lasers could also provide the next generation of parallel optical interconnects 
and optical information processing~\cite{Michalzik:2013,Ciftcioglu:2012}. Given the wide use of semiconductor lasers, improving their performance 
would have a huge impact. 

\begin{figure}[ht]
\resizebox{8.7cm}{!}{\includegraphics{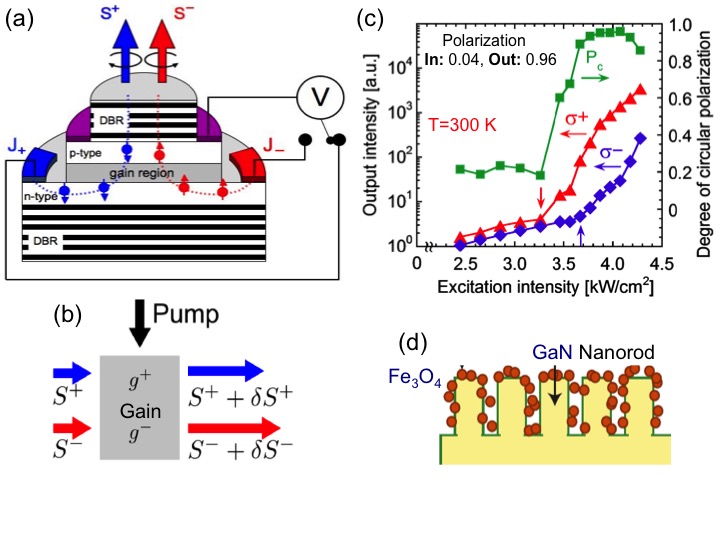}}
\vspace{-1.2truecm}
\caption{\footnotesize{%
(a) Spin laser with electrical spin injection~\cite{Sinova2012:NM}. The resonant cavity is made of the distributed Bragg reflectors (DBR). 
(b) Gain region: the photon density increases as it passes across the region, depending on the helicity, $S^{\pm}$~\cite{FariaJunior2015:PRB}.
(c) Output intensities ($\sigma_\pm$) and circular light polarization $P_c$ shown spin amplification in an optically injected 
GaAs-based spin laser~\cite{Iba2011:APL}. 
(d) Integrating Fe$_3$O$_4$ nanomagnets with the  gain region (GaN nanorods)~\cite{Chen2014:NN}.
Adapted with permission from (a) Ref.~\cite{Sinova2012:NM}, (b) from Ref.~\cite{FariaJunior2015:PRB}, (c) from Ref.~\cite{Iba2011:APL},
(d) from Ref.~\cite{Chen2014:NN}.
}}
\label{fig:VCSEL}
\end{figure}

Adding spin-polarized carriers in semiconductor lasers provides a new class of 
devices--spin lasers~\cite{Rudolph2003:APL,Holub2007:PRL,Hovel2008:APL,Gerhardt2011:APL,Frougier2013:APL} 
Their operation can be understood through transfer 
of angular momentum, the injection of spin-polarized carriers leads to the emission of circularly polarized light, depicted
in Fig.~\ref{fig:VCSEL}(a) for a so-called vertical cavity surface emitting laser (VCSEL). 
The ability to independently modulate the optical 
polarization and intensity in spin lasers leads to new operation regimes. As compared to their conventional (spin-unpolarized) counterparts, 
spin-lasers offer improved lasing threshold reduction~\cite{Rudolph2003:APL,Holub2007:PRL,Hovel2008:APL,Lee2014:APL} 
enhanced bandwidth~\cite{Lee2010:APL,Lee2012:PRB} reduced parasitic frequency modulation (chirp)~\cite{Boeris2012:APL} 
and error rates in digital operation~\cite{Wasner2015:APL}.

Conventional and spin lasers share three main elements: the gain region, the resonant cavity and the pump that injects (optically or electrically) carriers. The key effect of the gain region, typically quantum dot or quantum well, is producing a stimulated emission and coherent light that makes the laser such a unique light source. As shown in Fig.~\ref{fig:VCSEL}(b) for the schematic of the optical gain, in spin lasers the increase in photo density $\delta S$, 
depends on the helicity of light, $g^+ \neq g^-$. With their strongly nonlinear operation, spin lasers are efficient spin amplifiers: A small polarization of the injected carriers 
can lead to a nearly  complete polarization of the emitted light shown in Fig.~\ref{fig:VCSEL}(c), between the two lasing thresholds (vertical arrows)~\cite{Iba2011:APL}.

For practical applications of lasers their electrical pumping is most suitable. Some of the resulting challenges 
for electrically-operated spin lasers can be inferred from their device geometry depicted in Fig.~\ref{fig:VCSEL}(a). To achieve population inversion 
for lasing, a large carrier density is needed which also leads to shorter spin relaxation times and thus a shorter spin diffusion length. In typical spin 
lasers a spatial  separation between the spin injectors (blue/red  magnetic contacts) and the gain region of  several $\mu$m exceeds the spin diffusion
 length resulting in the carrier spin polarization negligible at room temperature. Bringing Fe$_3$O$_4$ nanomagnets  next to the gain region consisting 
 of GaN nanorods, as shown in Fig.~\ref{fig:VCSEL}(d), overcomes that limitation and led to the first electrically-controlled spin laser at room temperature~\cite{Chen2014:NN}.

\begin{figure}[ht]
\resizebox{8.6cm}{!}{\includegraphics{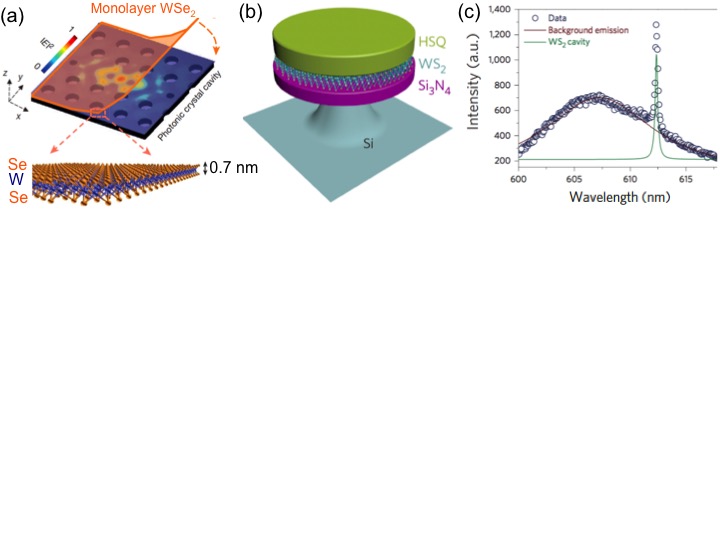}}
\vspace{-4.2cm}
\caption{\footnotesize{%
(a) Lasers with a monolayer TMD gain region. (a) WSe$_2$ with a photonic crystal~\cite{Wu2015:N}. (b) and (c) WS$_2$ microdisk 
excitonic laser~\cite{Ye2015:NP}. Photoluminescence with the narrow line characteristic for lasing. Adapted with permission from 
(a) Ref.~\cite{Wu2015:N}, (b) and (c) from Ref.~\cite{Ye2015:NP}.
}}
\label{fig:TMDLaser}
\end{figure}

With the integration of magnetic regions in spin lasers, magnetic proximity effects could be employed as electrically-tunable sources of spin-polarized carriers~\cite{Lazic2016:PRB} 
as well as to overcome the need for an applied magnetic field~\cite{Zutic2014:NN} in Ref.~\cite{Chen2014:NN} 
relying on paramagnetic nanomagnets. 
Rather than just implementing spin injection into III-V conventional semiconductors~\cite{Frougier2015:OE}, by placing F close to 
the gain region based on ML TMDs, the role of magnetic effects could be particularly pronounced. The feasibility of the proposed ML TMD-based 
spin lasers~\cite{Lee2014:APL} 
with desirable spin-dependent properties has been recently supported by the experimental demonstration of 
lasing in similar structures, shown in Fig.~\ref{fig:TMDLaser} which enable a very low lasing threshold~\cite{Wu2015:N, Ye2015:NP}. 
Vertical device geometries for lasers, as depicted in Fig.~21(a), could take 
advantage of TMD-based heterojunctions which for 
vertical stacking display improved properties, as compared to their lateral counterparts~\cite{Miao2017:ACSN}.

We propose that F region next to the ML TMD could be used to transform the excitons and thus enable a tunable operation of 
spin lasers by changing  the direction of {\bf M}, as suggested in Ref.~\cite{Scharf2017:PRL} [see also Figs.~9(f) and 9(g)]. 
Recent advances in vdW materials demonstrate a 2D ferromagnetism in 
a ML~\cite{Gong2017:N,Huang2017:N,Lin2015:JMCC,Liu2017:NPJ2DMA,Bonilla2018:NN} (see Sec.~3.2). 
These ferromagnets could open new directions for spin lasers with an atomically-thin gain region.   
Desirable properties of 2D vdW ferromagnets for vertical spin lasers, such as the perpendicular magnetic 
anisotropy (to remove the need for an applied magnetic field~\cite{FariaJunior2015:PRB}), 
room temperature and gate-controlled magnetism, and have already been demonstrated.  These
same properties are also valuable for many spintronic applications~\cite{Zutic2004:RMP,Tsymbal:2011}.
We expect that a future research will focus on dynamical response of 2D vdW ferromagnets and explore 
methods for their fast and low-energy switching. 

\section{6. Conclusions and Outlook}  
In this review we have explored a paradigm change in which proximity effects, 
commonly viewed as curious, but disjoint phenomena,  are instead 
considered as a versatile platform to transform a wide class of materials. With the advances
in heterostructures of reduced dimensions and improved interfacial quality we 
expect that the importance of proximity effects, despite their short characteristic length,
will only continue to grow. This trend is exemplified by van der Waals heterostructures in which
their constituent monolayers display the dominance of interfacial over bulk behavior, providing 
an ideal setting to test and tailor proximitized materials. 

Considering a steadily increasing number of these van der Waals 
materials~\cite{Duong2017:ACSN,Zhou2018:N,Mounet2018:NN} that are themselves ferromagnets, 
antiferromagnets, superconductors, or have a strong spin-orbit coupling, it is possible to now consider 
previously unexplored implications of  proximity effects in atomically-thin materials, including  
nontrivial topological properties. In fact, a surprising behavior, absent in the constituent materials, is already manifested 
in simple systems. A change in the stacking orientation between graphene and an insulator h-BN yields 
topological currents~\cite{Gorbachev2014:S}  while in a magnetic field graphene/h-BN heterostructures reveal a 
fractional quantum Hall effect with a peculiar fractal spectrum of a Hofstadter's butterfly~\cite{Dean2013:N,Hunt2013:S}. 
Remarkably, even a change in the stacking orientation between the two graphene layers can lead to striking results:
from the onset of superconductivity to the strongly-correlated insulator~\cite{Cao2018:Na,Cao2018:Nb}. While 
this approach deviates from the common picture of proximity effects which assumes different materials, 
together with the similar work on graphene/h-BN~\cite{Dean2013:N,Hunt2013:S} one can anticipate intensive 
efforts to explore how the twist angle between the neighboring van der Walls layers  and the resulting formation
of Moir\'e patters would alter materials properties in many other systems.

Some of the key opportunities in proximitized materials, both in their normal and superconducting state, 
rely on the interplay of multiple proximity effects. With its large conduction band exchange splitting and 
a large magnetic moment, the ferromagnetic insulator EuS, despite its low Curie temperature of $\sim16$ K was 
a common choice to implement magnetic proximity effects. Remarkably, a recent work on EuS heterostructures 
with topological insulators provides support for ferromagnetic ordering at room temperature~\cite{Katmis2016:N}. 
Even though the full explanation of this observation is yet to be obtained, it seems that the strong spin-orbit coupling of 
a topological insulator could be responsible for strengthening of the magnetic order in EuS. At the molecular scale 
there are also examples showing that magnetism can be enhanced with nonmagnetic molecules~\cite{Raman2013:N}. 

While experimental reports of skyrmions have spanned a large class of materials and systems: from lattices in quantum Hall effect, Bose-Einstein condensates, and polaritons, and to topological insulators and multiferroics~\cite{Schmeller1996:PRL,Schweikhard2004:PRL,Flayac2013:PRL,Seki2012:S,Yasuda2016:NP,Flayac2013:PRL}
the recent attention has mostly focused on magnetic skyrmions, 
as a versatile building block for spin-based devices and even more complex topological states. Only a few bulk systems support 
stable skyrmions, typically limited to a narrow region of the temperature-magnetic field phase diagram and well below room temperature. It is therefore crucial to engineer proximity effects and interfaces to ensure their stability~\cite{Romming2013:S,Dupe2014:NC} 
and enable controlled creation and manipulation of individual skyrmions~\cite{Hsu2017:NN,Schott2017:NL}. 
The usual approach is to seek a large chiral Dzyaloshinskii-Moriya interaction~\cite{Tsymbal:2011}  
by using a layer of heavy metal atoms with strong spin-orbit coupling like Pd or Ir in contact with a 3$d$ ferromagnet like Fe. Surprisingly, 
a relatively strong Dzyaloshinskii-Moriya interaction can also be realized by proximity-induced spin-orbit coupling in systems 
without heavy elements, such as graphene-covered ultrathin Co or Ni films~\cite{Yang2017:P}. It was also shown that a coupled 
pair of skyrmions of opposite chiralities can be stabilized in a magnetic bilayer,  where dipole coupling allows the skyrmion 
pairs to be stabilized without the need for a very large  
Dzyaloshinskii-Moriya interaction~\cite{Hrabec2017:NC}. 

Requirements to realize Majorana bound states for implementing fault-tolerant quantum computing call for elusive spin-triplet 
superconductivity~\cite{Aasen2016:PRX,DasSarma2015:QI,Kitaev2003:AP}, in contrast to common superconductors made 
of spin-singlet Cooper pairs. Remarkably, a lack of naturally occurring triplet superconductors is overcome by a careful design 
of proximity effects where the superconductivity is induced in a semiconductor host with strong spin-orbit coupling removing 
the usual pairing between spin-up and spin-down electrons and making their spin-triplet pairing preferable. Placing a nearby 
array of ferromagnets can even remove the need for a strong spin-orbit coupling. The resulting fringing fields themselves can 
induce effective spin-orbit coupling and control the formation of Majorana bounds 
states~\cite{Klinovaja2012:PRL,Fatin2016:PRL,Matos-Abiague2017:SSC}.  
An interplay between magnetic and superconducting proximity effects has already been extensively studied in superconducting spintronics. 
Superconductor/ferromagnet junctions with noncollinear magnetization or spin-orbit coupling
support the formation of long range spin-triplet proximity effects and the control of pure spin currents~\cite{Keizer2006:N,Hogl2015:PRL,Eschrig2015:RPP,Linder2015:NP,Halterman2013:PRL,%
Bergeret2005:RMP,Buzdin2005:RMP}. 
We can expect that many normal state proximity effects discussed in this review will also lead to intriguing superconducting counterparts.

\acknowledgments
We thank J. M. D. Coey, M. Eschrig, J. Fabian, N. C. Gerhardt,  J. Han, W. Han, E. Hankiewicz, R. Kawakami, P. Lazi\'c, A. Petrou, 
J. Shi, V. Svetli\v{c}i\'c, S. Valenzuela, %
G. Xu, J. Xu, H. Zeng, and T. Zhou for valuable discussions. 
I. \v{Z}. was supported by the Department of Energy, Basic Energy Sciences (Grant No. {DE-SC0004890}), 
the Office of Naval Research (Grant No. 000141712793), and the National Science Foundation (Grant. No. 
ECCS-1508873).
A. M-A. was supported by by the Department of Energy, Basic Energy Sciences (Grant No. {DE-SC0004890})
and the Office of Naval Research (Grant No.  000141712793).
Computational work was supported by the UB Center for Computational Research and the Unity Through Knowledge Fund
(Contract No. 22/15). 
B. S. was supported by the German Science Foundation (DFG) (Grant No.  
SFB 1170 ``ToCoTronics") and the ENB Graduate School on Topological Insulators.
H.D. was supported by the Department of Energy, Basic Energy Sciences (Grant No. {DE-SC0014349}), the National Science 
Foundation (Grant No. DMR-1503601), and the Defense Threat Reduction Agency (Grant No. HDTRA1-13- 1-0013). 
K. B. was supported by the National Science Foundation  (Grant No. DMR-1609776 and the Nebraska MRSEC, 
Grant No. DMR-1420645), and the Nanoelectronics Research Corporation (NERC), a wholly-owned subsidiary of the 
Semiconductor Research Corporation (SRC), through the Center for Nanoferroic Devices (CNFD), a SRC-NRI 
Nanoelectronics Research Initiative Center (Task IDs 2398.001 and 2398.003).

%
\end{document}